\documentclass[10pt,journal,compsoc]{IEEEtran}
\ifCLASSOPTIONcompsoc
  \usepackage[nocompress]{cite}
\else
  \usepackage{cite}
\fi

\usepackage{hyperref}
\usepackage{url}
\usepackage{color}

    \usepackage[pdftex]{graphicx}
   \DeclareGraphicsExtensions{.pdf,.jpeg,.png}

\usepackage{mathptmx}
\usepackage{graphicx}
\usepackage{times}

\usepackage{textcomp}
\usepackage{url}
\usepackage{verbatim}
\usepackage{cite}

\usepackage{balance}

\usepackage{amsmath,amsfonts}
\usepackage{setspace}
\usepackage{color, colortbl}
\usepackage{pdfpages}
\usepackage[linesnumbered,ruled]{algorithm2e}

\usepackage{balance}

\usepackage{mathptmx}                  
\usepackage{amssymb}
\usepackage{bbm}
\usepackage{wrapfig}
\usepackage{subcaption}
\usepackage{hyperref}
\newtheorem{definition}{Definition}

\newcommand{\myparagraph}[1]{\mbox{\ } \newline \noindent \textbf{#1}}
\renewcommand{\paragraph}[1]{\myparagraph{#1}}

\usepackage{tikz}

\begin{document}
%
%
\title{Hybrid Base Complex: Extract and Visualize Structure of Hex-dominant Meshes}



\author{Lei Si, Haowei Cao, Guoning Chen~\IEEEmembership{Senior Member, IEEE}
\IEEEcompsocitemizethanks{\IEEEcompsocthanksitem Lei Si, Haowei Cao, and Guoning Chen are with the University of Houston. E-mail: (lsi, hcao13)@uh.edu, gchen22@central.uh.edu.\protect
}
}

%
%

\markboth{IEEE TRANSACTIONS ON VISUALIZATION AND COMPUTER GRAPHICS,~Vol.~-, No.~-, -~2021}%
{Shell \MakeLowercase{\textit{et al.}}: Bare Demo of IEEEtran.cls for Computer Society Journals}
%



\IEEEtitleabstractindextext{%
\begin{abstract}
    Hex-dominant mesh generation has received significant attention in recent research due to its superior robustness compared to pure hex-mesh generation techniques. In this work, we introduce the first structure for analyzing hex-dominant meshes. This structure builds on the base complex of pure hex-meshes but incorporates the non-hex elements for a more comprehensive and complete representation. We provide its definition and describe its construction steps. Based on this structure, we present an extraction and categorization of sheets using advanced graph matching techniques to handle the non-hex elements. This enables us to develop an enhanced visual analysis of the structure for any hex-dominant meshes.We apply this structure-based visual analysis to compare hex-dominant meshes generated by different methods to study their advantages and disadvantages. This complements the standard quality metric based on the non-hex element percentage for hex-dominant meshes.
    Moreover, we propose a strategy to extract a cleaned (optimized) valence-based singularity graph wireframe to analyze the structure for both mesh and sheets. Our results demonstrate that the proposed hybrid base complex provides a coarse representation for mesh element, and the proposed valence singularity graph wireframe provides a better internal visualization of hex-dominant meshes.
\end{abstract}

\begin{IEEEkeywords}
Structure Analysis, Hex-dominant Meshes, Structure Visualization
\end{IEEEkeywords}}

\maketitle

\setlength{\abovedisplayskip}{2pt}
\setlength{\belowdisplayskip}{2pt}
\setlength{\abovedisplayshortskip}{2pt}
\setlength{\belowdisplayshortskip}{2pt}
\setlength{\belowcaptionskip}{3pt}
\setlength{\abovecaptionskip}{3pt}
\setlength{\textfloatsep}{3pt}
\setlength{\floatsep}{0pt}
\setlength{\intextsep}{2pt}

\IEEEdisplaynontitleabstractindextext

%
\IEEEpeerreviewmaketitle

\IEEEraisesectionheading{\section{Introduction}\label{sec:intro}}

\begin{figure*}
  \includegraphics[width=\linewidth]{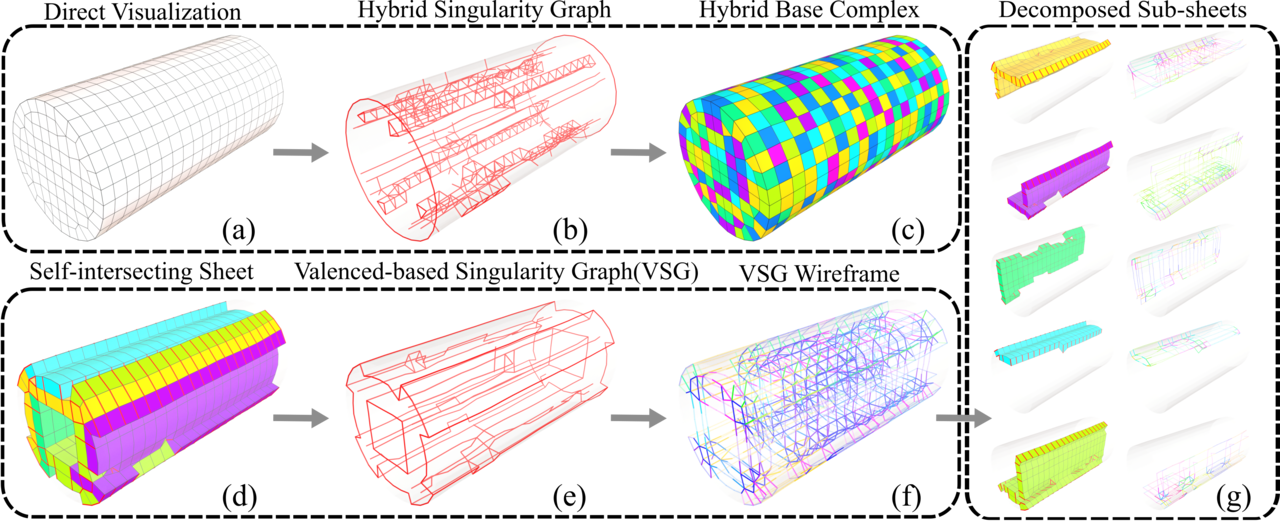}
  \centering
  \caption{Illustration of the pipeline of our method.  
  Given a hex-dominant mesh (a), we first extract the hybrid singularity graph (b) that includes conventional singularities and the edges of all non-hex cells. From this hybrid singularity structure, we construct a hybrid base complex (c). Different colored blocks correspond to individual hex and non-hex components. From the hybrid base complex, individual hexahedral sheets are extracted that encode the substructure configurations. (d) shows a self-intersecting sheet of this mesh. To better study the internal configurations of this sheet, we extract a valence-based singularity graph, denoted by $VSG$, which contains all irregular edges of this sheet. Note that $VSG$ is more complex than the one in (b), as it includes all the boundary sharp features of the sheet. From $VSG$, we extract a connected network, called $VSG$ wireframe that encodes the non-hex configurations and their propagation within the sheet (f). They are important for the subsequent non-hex element removal. This self-intersecting sheet can be further decomposed into simpler subsheets for a detailed analysis (g).
  }
  \label{fig:teaser}
\end{figure*}

\IEEEPARstart{V}{olumetric} meshes consisting entirely of hexahedra tend to be more efficient and easier to work with when performing finite element analysis (FEA) due to the desired numerical properties of hexahedra \cite{10.1145/3313797, tadepalli2011comparison, schneider2022large}.
However, the automatic generation of high-quality, geometry-conforming, all hexahedral meshes for arbitrary models has been one of the most challenging meshing problems for over three decades \cite{beaufort2021hex, 10.1145/3197517.3201344}. 
As a compromise, hexahedral (hex-) dominant meshing is considered to improve robustness and overall mesh quality by introducing a small ratio of generic polyhedral elements \cite{pietroni2022hex}. In the last decade, most of the research in this direction has focused on achieving the highest proportion of hexahedra in the meshes produced \cite{BERNARD201678, livesu2020loopycuts, gao2017robust, bukenberger2022most}. 
Nevertheless, generating all-hex meshes from these hex-dominant meshes remains an open problem, despite being highly sought-after by domain experts.

Due to varying configurations of the non-hex elements (or cells) (e.g., with different numbers of vertices, edges, and faces; different orientations; non-conformality at interfaces), reducing (or eliminating) such elements from a hex-dominant mesh is non-trivial. Often, changes in the connectivity of these non-hex elements will propagate throughout the mesh; therefore, knowing how these changes propagate is crucial for the design of operations to robustly remove non-hex elements. To achieve this, a global structure of hex-dominant meshes is desired that incorporates both hex and non-hex elements as well as describes the larger organization of elements (similar to the hexahedral blocks in all-hex mesh \cite{gao2015hexahedral,xu2018hexahedral}). With such a global structure and its partitioning of the mesh into larger components, the quality of a hex-dominant mesh can be evaluated globally (i.e., the fewer the overall large components and the fewer the non-hex components, the better the quality of the hex-dominant mesh). 
However, to our knowledge, there is no definition of such a global structure for hexahedral dominant meshes. 

To fill this gap, we introduce a first complete structure for hex-dominant meshes that extends the \emph{base complex} of all-hex meshes \cite{gao2015hexahedral,xu2018hexahedral} by incorporating non-hex elements. 
In all-hex meshes, a base complex is a partition of the mesh into a set of large hexahedral blocks called base complex components. 

\begin{figure}
     \centering
     \begin{subfigure}[b]{0.49\columnwidth}
         \centering
         \includegraphics[width=0.95\columnwidth]{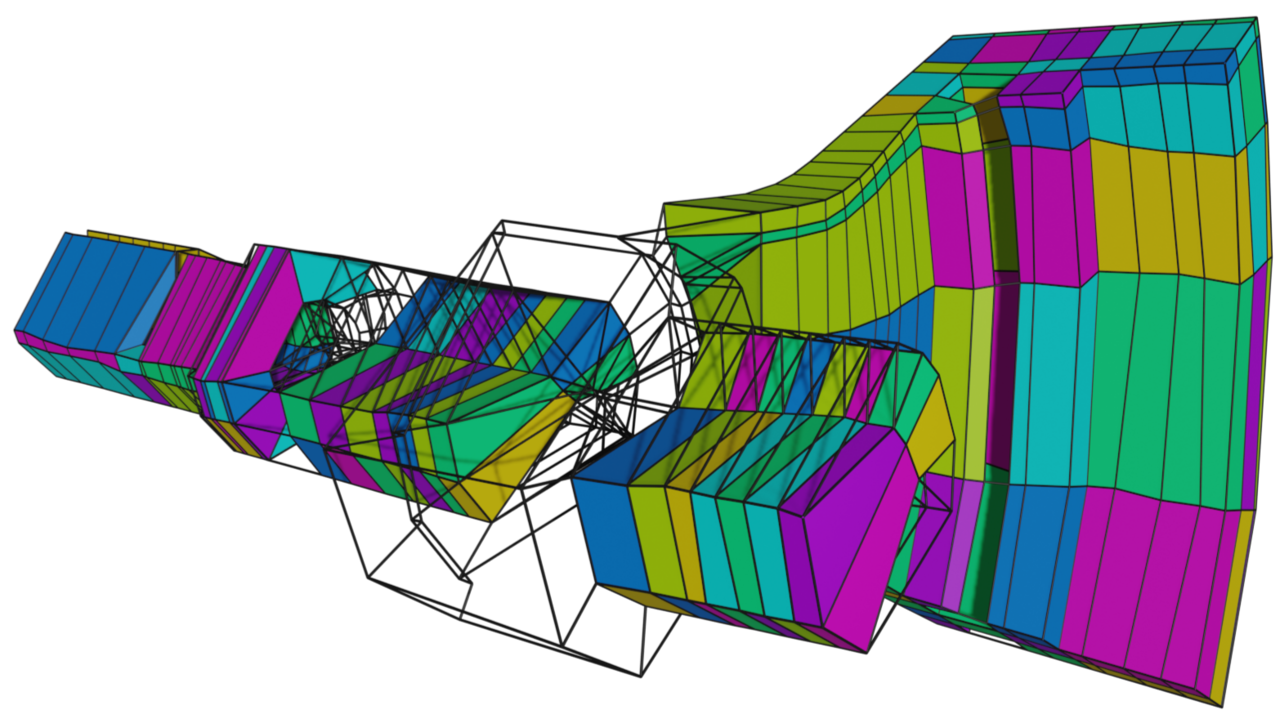}
         \caption{Base Complex}
         \label{fig:BC_hex_only}
     \end{subfigure}
     \begin{subfigure}[b]{0.49\columnwidth}
         \centering
         \includegraphics[width=0.95\columnwidth]{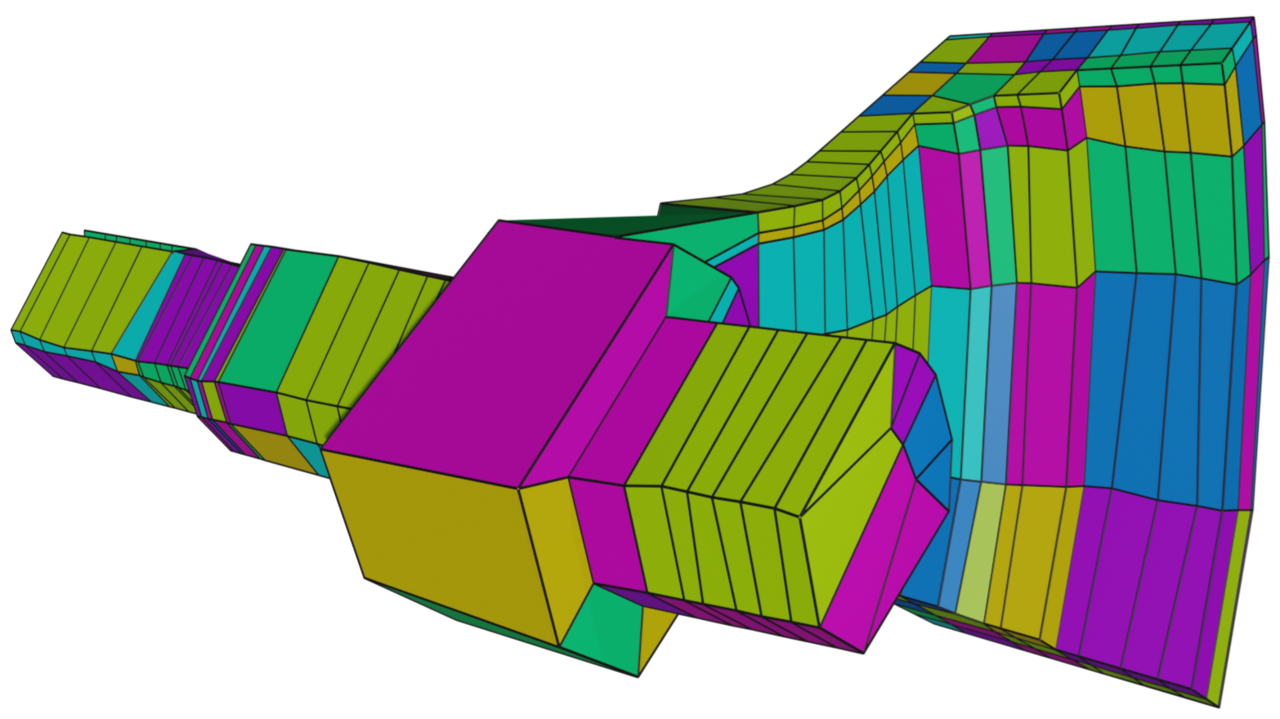}
         \caption{Hybrid Base Complex}
         \label{fig:HBC_nugear}
     \end{subfigure}
        \caption{Conventional base complex with gaps (a) vs the hybrid base complex with non-hex component reduction (b) for a hex-dominant mesh.}
        \label{fig:BSvsHBS}
\end{figure}

For hex-dominant meshes, a base complex (with gaps) can still be extracted from the portions of the volume that are comprised of hexahedra (\autoref{fig:BSvsHBS}a). To achieve full coverage of the entire volume, we incorporate edges involved with non-hex elements into the singularity structure and define a \emph{hybrid singularity structure} (\autoref{fig:teaser}b), from which a \emph{hybrid base complex} can be constructed that is comprised of not only hexahedral components but also non-hex components (\autoref{fig:teaser}c and \autoref{fig:BSvsHBS}b).

The hybrid base complex can depict the complexity of the corresponding hex-dominant mesh, such as the alignment of the orientation of the hexahedral elements, which is often overlooked when evaluating the quality of hex-dominant meshes. 
However, visualizing the hybrid base complex using colored blocks that correspond to individual base complex components usually leads to occlusion and clutter, which does not provide an effective depiction of the (inner) configurations of the structures (see \autoref{fig:teaser}c). To enable a detailed analysis of hex-dominant mesh structures and their comparison, we extract substructures, called \emph{sheets}, from the hybrid base complex. Our sheets adapt the hexahedral sheets \cite{xu2018hexahedral} but are augmented to incorporate the information of the adjacent non-hex elements. These adjacent non-hex elements serve as corridors to connect nearby sheets, which indicates the potential strategy of removing them. Thus, they are of particular interest to meshing practitioners and simplification algorithms. 

Although sheets reduce the visual complexity, displaying a sheet with all involved blocks and adjacent non-hex elements may still lead to visual clutter (see \autoref{fig:teaser}d). To address this, we classify sheets into different types and propose a strategy to decompose complex sheets into simple ones (\autoref{fig:teaser}g). Furthermore, we introduce the \emph{valence-based singularity graph} (\autoref{fig:teaser}e) and its derived wireframe (\autoref{fig:teaser} f) to intuitively describe irregular configurations, including non-hex configurations and how they propagate across the volume and complicate the structures. 
We incorporate our hybrid base complex construction, sheet extraction, and visual representation extraction and simplification process into a first visual analysis framework for the effective analysis and comparison of different hex-dominant meshes.

In summary, our work makes the following contributions.

\begin{itemize}
    \item We introduce a first global structure encompassing the non-hex components for hex-dominant meshes. 
    \item We extend the concept of sheets in the base complex to the hybrid base complex and introduce an effective algorithm to extract them using the graph matching techniques \cite{gibbons1985algorithmic, west2001introduction}. The configuration of these sheets can be utilized for the evaluation of structural complexity of hex dominant meshes, which was not possible previously.
    \item We introduce a first set of visualization techniques for the constructed hybrid base complex and its sheets (and subsheets) to support an intuitive and effective qualitative evaluation of hex-dominant mesh structure.
    \item We propose a valence-based singularity graph and its corresponding connected wireframe to exploit the internal mesh structure quality for both the entire mesh and its sheets.
\end{itemize}

We have applied our hybrid base complex extraction and visualization to analyze and compare the complexity and characteristics of a set of hex-dominant meshes produced by several state-of-the-art techniques. Our results show that our hybrid base complex and its sheets can effectively reveal the difference among different hex-dominant meshes. They not only show the complex configurations that are caused by a few simple non-hex elements but also intuitively convey the difficulty of removing these elements.
The proposed hybrid base complex and its extended sheets set the foundation for a quantitative study of the characteristics and quality of the structure of any hex-dominant meshes similar to the work in \cite{xu2018hexahedral} for pure hex-meshes.

We attach all results in the additional material and will release the source code for a reference implementation of the proposed framework on Github.

\section{Related Works}

\subsection{Hexahedral-dominant Meshing}
\label{sec:hexdom-meshing}

Hex-dominant meshing techniques closely relate to all-hex mesh generation. There are numerous efforts in this regard, like sweeping \cite{Gao2015TVCG}, octree-based \cite{gao2019feature, pitzalis2021generalized}, polycubes mapping \cite{HexmeshSGP2011,livesu2013polycut, huang2014,Fang2016AMU, guo2020cut}, and frame field based methods \cite{Huang2011,NieserSGP11, Li2012, Jiang2013} for all-hex mesh generation. Nonetheless, automatic generation of high-quality and feature-aligned all-hex meshes for arbitrary models remains the ``holy grail'' problem for the meshing and geometric modeling communities. While the recent mapping and parameterization techniques can produce high-quality all-hex meshes, they often fail for complex models due to the lack of guarantee in numerical approximation \cite{blacker2001automated, pietroni2022hex}. This is particularly true for some complex models described in \cite{ray2018hex}.

Consequently, hex-dominant meshing is getting increased attention due to its robustness when handling complex models. 
The goal of hex-dominant meshing is to tile as many hexahedral elements in the volume as possible while accepting a small number of non-hex elements~\cite{pietroni2022hex}.
To achieve this goal, Meshkat and Talmor \cite{meshkat2000generating} produced hex-dominant meshes by aggregating neighboring tetrahedrons to assemble hexahedral cells. Similarly, Owen \cite{owen2001hex} introduced H-morph to gradually transform a tetrahedral mesh into a hex-dominant mesh starting from the boundary that was already converted into a quadrilateral mesh. The non-hex elements in these methods are usually simple (e.g., prism, pyramid, or tetrahedra).

Yamakawa and Shimada \cite{yamakawa2002hexhoop} introduced HEXHOOP to convert a hex-dominant mesh to an all-hex mesh. The proposed method is capable of automatically converting a hex-dominant mesh to an all-hex mesh by subdividing a prism/pyramid/tetrahedral element into a set of smaller hexahedral elements, while ensuring topological conformity with neighboring elements. However, it cannot handle other more complex non-hex element types (e.g., polyhedra with arbitrary numbers of faces).
Pellerin et al.\cite{pellerin2018identifying} used a vertex-based strategy to combine tetrahedra into hexahedra. However, there is no guarantee that all tetrahedra can be combined into hexahedra.

While the computation of $L_p$ centroidal Voronoi tessellation \cite{levy2010p} and boundary conformal 3D cross field \cite{Huang2011} can be used to produce hex-dominant meshes, 
Sokolov et al.\cite{sokolov2016hexahedral} extended the periodic global parameterization for surfaces \cite{ray2006periodic} to 3D and proposed the first field-aligned parameterization method to guide the agglomeration of tetrahedra to produce hex-dominant meshes. This method was later improved by Gao et al.\cite{gao2017robust} to address the non-conformality in the meshes produced by the former method. However, the non-hex elements produced by this method can be arbitrarily complex (e.g., with up to 30 faces). 
Livesu et al. \cite{livesu2020loopycuts} proposed a fully automatic algorithm to produce hex-dominant meshes by mimicking manual block decomposition. 
Yu et al. \cite{yu2022hexdom} created hex-dominant meshes by employing a multi-model polycube-based algorithm that requires manual intervention. 
The recent approach proposed by Bukenberger et al. \cite{bukenberger2022most} generates at-most-hex meshes based on a 3D Lloyd relaxation under the $L_\infty$ norm for a harmonious hexahedral cell layout. In the resulting meshes, no cell has more than six faces, and no boundary face has more than four sides. Despite that different methods produce hex-meshes with non-hex elements of different types, our structural representation can handle all of them.

\subsection{Hex-Mesh Structure and its Visualization}
\label{sec:hexstructure}

Our proposed structure is closely related to the \emph{base complex} of all-hex meshes \cite{gao2015hexahedral,gao2017b}. 
To construct the base complex for a hex-mesh, the irregular edges whose numbers of adjacent hex elements are not 4 in the interior or not 2 on the boundary are first extracted. The connected irregular edges form \emph{singularities}. From singularities, separation surfaces are traced out, which partitions the mesh into hexahedral blocks. This partitioning is the base complex. 
The complexity of a base complex impacts the quality of the corresponding mesh \cite{gao2017b} and the subsequent fitting of high-order basis functions for PDE solving \cite{Gao2015TVCG,IMR19abstract}. In general, the complexity of the base complex is measured by its number of hexahedral blocks. A smaller number of blocks is preferred as the corresponding mesh is considered (semi-) structured. Singularity alignment technique \cite{gao2015hexahedral} and sheet removal method \cite{gao2017robust} have been proposed to simplify base complexes.  An alternative to the conventional base complex is the 3D motorcycle complexes introduced by Br{\"u}ckler et al. \cite{10.1111:cgf.14470}, which generalizes the 2D motorcycle graphs \cite{eppstein2008motorcycle} for quad meshes. Different from a base complex, a motorcycle complex allows the existence of T-junctions, leading to the generation of T-meshes that need to be converted to all-hex meshes \cite{bruckler2022volume}. It is worth noting that Schertler et al. \cite{schertler2018generalized} utilized the 2D motorcycle graph as a structure representation for quad-dominant meshes. In summary, there is no existing work on the study of the structure in hex-dominant meshes. The proposed structure in this work fills such a gap.

There do not exist many techniques for the visualization of hex- and hex-dominant meshes, let alone their respective structures. 
Bracci et al. \cite{bracci2019hexalab} presented an online tool, called hexaLab.net, to offer the publication-quality rendering of all-hex meshes and support simple element quality inspection using various cutting and filtering strategies. 
To address the occluding and cluttering issue for the inspection of hexahedral meshes, a focus+context volume rendering technique \cite{neuhauser2021interactive} has been introduced that assigns high opacity values to regions with poor quality elements.
Lei and Chen \cite{LeiVis23} introduced a visual analysis system for the study of the quality of all hex meshes (as well as quad meshes). Their system reveals regions with small, poor-quality elements that other methods cannot. None of these methods visualizes the structure of the meshes.

To visually analyze base complexes, Xu et al. \cite{xu2018hexahedral} introduced a multi-level decomposition that subdivides a base complex into a set of sheets, from which a set of \emph{main sheets} that best represent the base complex is selected for visualization. To reduce visual clutter, surface sheets are reduced to a set of curves following the two parameterization directions of the surface sheet. While effective for the visual analysis of base complexes and their comparison, that technique cannot be directly applied to the proposed structure for hex-dominant meshes. In addition, the focus of the hex-dominant mesh structure is on how different non-hex elements complicate the structure configurations, which requires new representations and a different set of visualization methods to reveal. The presented visualizations address this need.

\section{Generalize base complex for Hex-dominant meshes}
\label{sec:nonhexstructure}

In this section, we first define a new hybrid singularity structure for a hex-dominant mesh (Section \ref{sec:hybridstructure}). From this new structure, we further introduce a hybrid base complex for hex-dominant meshes (Section \ref{sec:hybridbasecomplex}) that extends the conventional base complex for all-hex meshes by borrowing some concepts from the recently introduced motorcycle complex \cite{10.1111:cgf.14470}. 

\subsection{Hex-dominant Mesh}
\label{sec:hexdominantmeshdef}

Hex-dominant mesh is a volumetric mesh that contains a small number of non-hex cells.
Mathematically, a volumetric mesh can be represented as a 3D network $G = (V,E,F,C)$ where $V$ is a set of vertices, $E$ is a set of edges, $F$ is a set of faces, and $C$ is a set of volume cells given a volume $\Omega$ with closed boundary $\partial \Omega$. 

Throughout this paper, the set of volume (3D) cells $C$ contains two subsets $(H, \hat{H})$ where $H$ is a set of hex cells and $\hat{H}$ is a set of non-hex cells. The valence of a vertex, $n(v_i)$, or an edge, $n(e_i)$, is defined as the number of adjacent cells.

\subsection{Hybrid Singularity Graphs in Hex-dominant Meshes}
\label{sec:hybridstructure}

The conventional singularity structure is not well-defined at non-hex cells, thus it cannot cover the entire space of a hex-dominant mesh $G$. To address that, we introduce a \emph{hybrid singularity structure}, which is a graph denoted by $G_S = (V_S \cup V_{\hat{H}}, E_S \cup E_{\hat{H}})$ (\autoref{fig:teaser}b). $E_S$ is the set of singularities, each of which is a set of connected edges (adjacent to pure hex cells in $H$) that have the same irregular valence (i.e., not equal to 4 in the interior or not equal to 2 on the boundary). $V_S$ is the set of singular vertices that are usually the two end vertices of a singularity. In other words, $E_S$ and $V_S$ are the conventional singularities and singular vertices as defined in the previous work \cite{xu2018hexahedral}. 

$E_{\hat{H}}$ is the set of non-hex edges, each of which is incident to at least one non-hex cell (in $\hat{H}$). For consistency, we refer to this set of edges as \emph{ pseudo-singularity}. Although non-hex edges can connect with each other, they often do not have the same valence, thus, for simplicity, we treat each non-hex edge as a pseudo-singularity. Note that a pseudo-singularity cannot be part of a singularity and vice versa. $V_{\hat{H}}$ is comprised of vertices that mark the two endpoints of pseudo-singularity. Although these vertices can be regular, they disrupt the mesh global structure. Therefore, when constructing singularity or pseudo-singularity, these edges must be terminated at these vertices. We refer to these vertices as \emph{pseudo singular vertices}. Note that singularities and pseudo-singularities may meet at a vertex, which is a pseudo singular vertex, not a singular vertex. 

\subsection{Hybrid Base Complex of Hex-Dominant Mesh}
\label{sec:hybridbasecomplex}

\textbf{Hybrid base complex: } 
Given the hybrid singularity structure $G_S$ of a hex-dominant mesh $G$, a hybrid base complex can be defined and constructed by extending the conventional base complex for all-hex meshes \cite{gao2015hexahedral,xu2018hexahedral}. 
Specifically, for each singularity or pseudo-singularity with valence $n$ in $G_S$, $n$ separation surfaces can be traced out that end at other singularities or pseudo-singularities or at the boundary $\partial \Omega$. 

These separation surfaces and their intersections partition the volume $\Omega$ into either topologically cube-like components, rings \cite{xu2018hexahedral}, or other non-hex components. Cube-like components are only specific to the components that contain six faces, and all faces are topologically quads. The ring-like components can be converted to cubes by cutting \cite{xu2018hexahedral}. The types of non-hex components depend on the types of non-hex elements in the input meshes, which can be arbitrary \cite{gao2017robust}.

This partitioning defines the \emph{hybrid base complex} $G_B$ of the hex-dominant mesh $G$ (\autoref{fig:teaser}d). Similar to the conventional base complex, $G_B$ provides a coarse tessellation of $\Omega$ (and an organization of the elements in $G$). 
We denote $G_B = (V_B, E_B, F_B, C_B)$, where $C_B$ is the set of base complex components. $C_{B} = (H_{B} \cup \hat{H}_B)$ with $H_B$ being hex-components and $\hat{H}_B$ representing non-hex components.
$F_B$ is the set of base complex faces that form the individual base complex components and can be quadrilateral or any non-quadrilateral polygons. $E_B$ are the edges of the individual base complex faces, and $V_B$ are the corners (or vertices) of the individual base complex components. 
Based on this definition, each non-hex cell in $G$ leads to a separate non-hex component.

\section{Sheets in Hybrid Base Complex}
\label{sec:sheets}

\begin{figure}
     \centering
     \begin{subfigure}[b]{0.58\columnwidth}
         \centering
         \includegraphics[width=0.85\columnwidth]{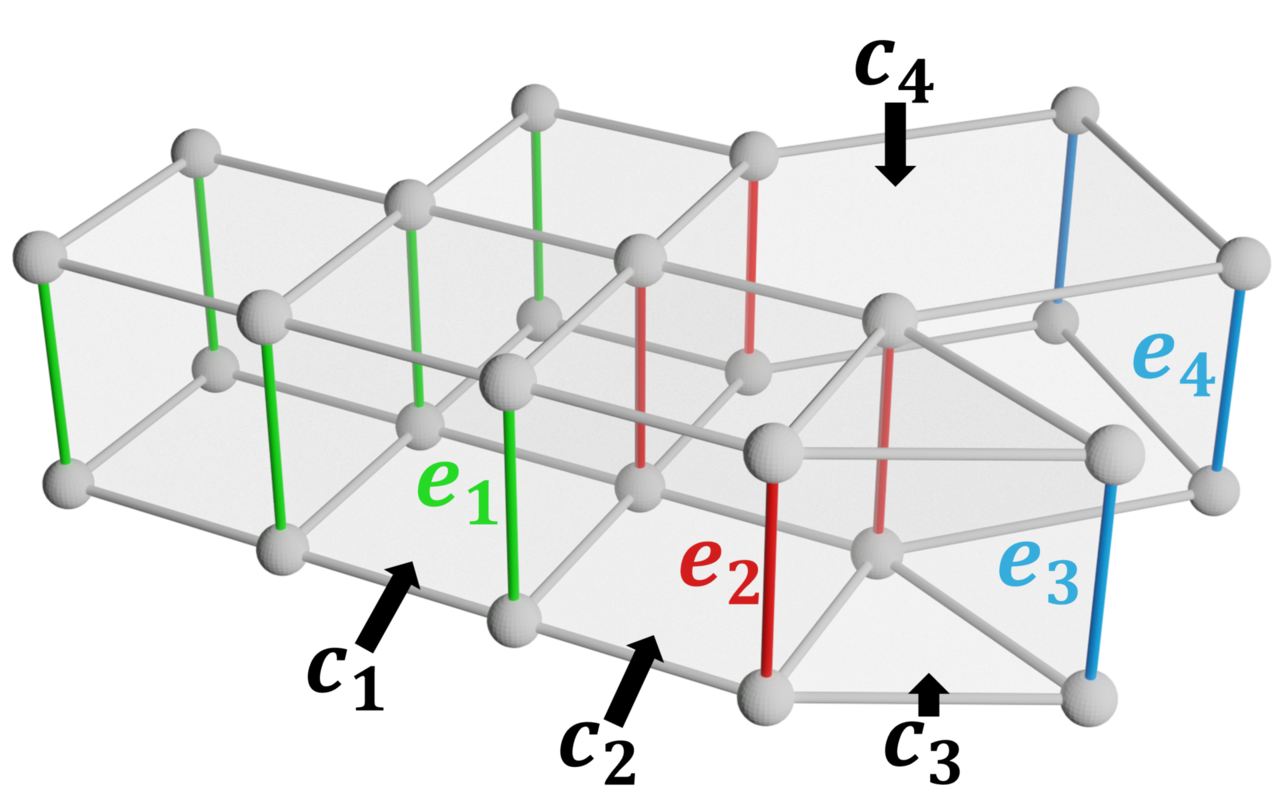}
         \caption{}
         \label{fig:sheet_example_1}
     \end{subfigure}
     \begin{subfigure}[b]{0.3\columnwidth}
         \centering
         \includegraphics[width=0.85\columnwidth]{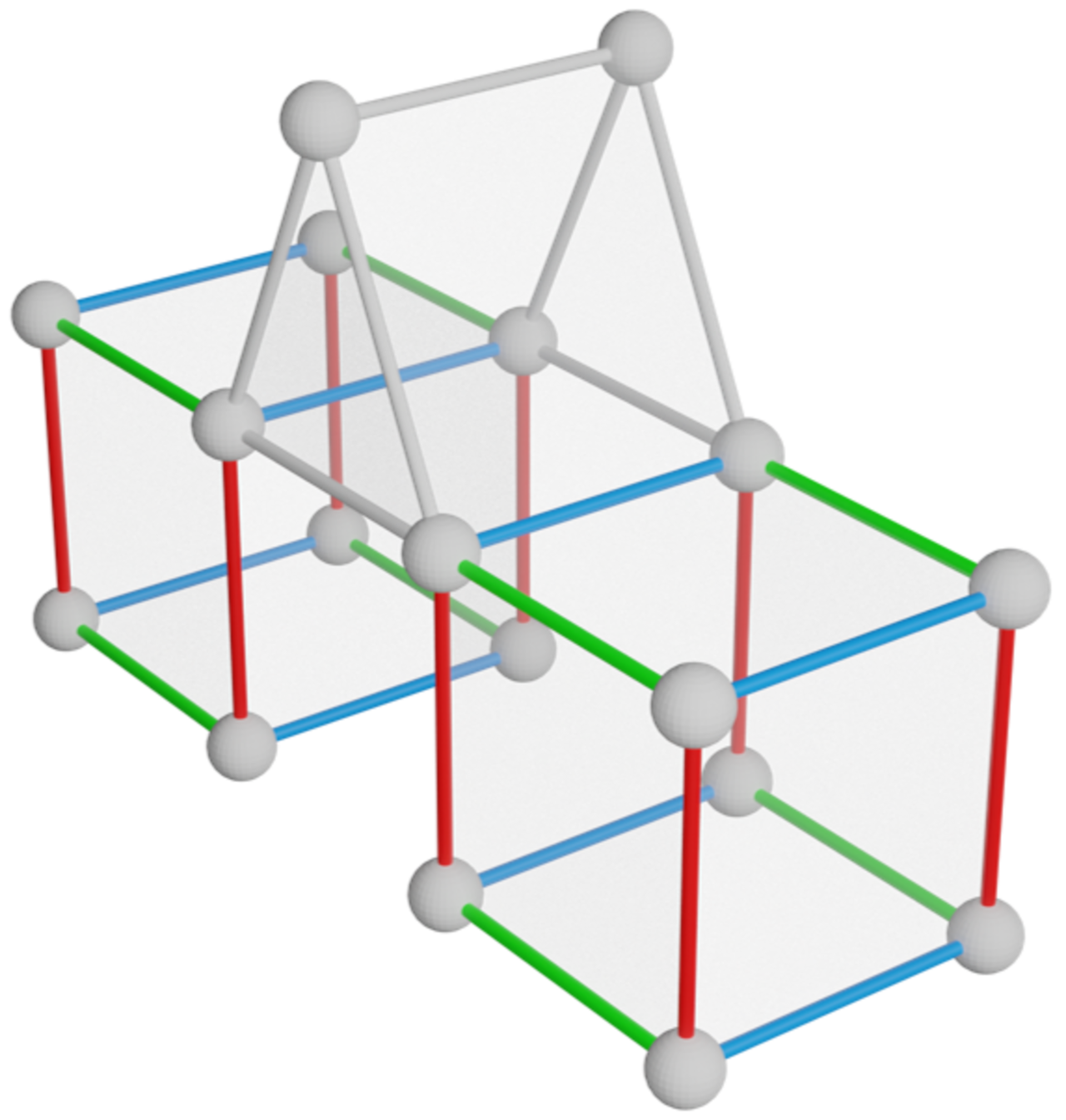}
         \caption{}
         \label{fig:sheet_break_by_non_hex}
     \end{subfigure}
        \caption{(a) a sheet is constructed by a set of parallel edges. (b) highlights a hexahedral block with its three orthogonal sets of parallel edges color-coded in red, green, and blue, respectively. The two hex blocks in (b) belong to two different sheets.}
        \label{fig:basic_sheet_example}
\end{figure}

To understand the detailed orientation configurations of the hybrid base complex, we decompose it into individual sheets. 
Given the hybrid base complex for a mesh $G$, a sheet $G_{SH} = (V_{SH}, E_{SH}, F_{SH}, C_{SH})$ comprises a set of \emph{hexahedral} (base complex) blocks identified by a set of parallel (base complex) edges $E_P$. 
Parallel edges are defined as follows:

\begin{definition}[parallel edges]
Edges $e_i$ and $e_j$ are considered parallel, denoted as $e_i \parallel e_j$, if and only if they belong to the same quad face but do not share any vertex.
\end{definition}

Two hex blocks belong to the same sheet if they are adjacent by a face and contain a set of parallel edges. For example, \autoref{fig:sheet_example_1}, $c_1, c_2$ are hex blocks with $e_1$ and $e_2$ being parallel, considering both $c_1$ and $c_2$ within the same sheet. 

Note that the concept of sheets above can be applied to the original mesh $G$ by grouping the connected hex cells (instead of blocks) through the corresponding parallel mesh edges.

\subsection{Sheet Construction}

We extract sheets using only hex blocks by following the method proposed in 
\cite{gao2017b}. The process starts with a hex block and extends to its connected hex blocks by tracing an edge and all its parallel edges. This construction continues until no other parallel edges can be found in the connected hex blocks.

Every hex block contains three sets of parallel edges oriented in three distinct directions. As illustrated in \autoref{fig:sheet_break_by_non_hex}, for each hex element, the red, green, and blue parallel edge sets propagate in orthogonal directions, leading to three orthogonal sheets. However, due to irregular configurations, a sheet may self-intersect. This means that, within a sheet, multiple parallel sets are traveled through a hex in different directions. A demonstration is shown in \autoref{fig:self_intersecting}. This configuration is important for our structural visualization and analysis, which we will discuss later.

After extracting sheets, we augment them with neighboring non-hex elements. 
This is because the above sheets only contain hex components. Adding non-hex elements can fill the gaps between those sheets, making it possible to study the propagation of the non-hex configurations in the volume (\autoref{sec:vsg}).
A non-hex element is considered adjacent to a sheet if the edges in a non-hex element are also in the parallel edge set of a sheet. As in \autoref{fig:sheet_example_1}, non-hex cells $c_3$ and $c_4$ are adjacent to the (green) sheet due to both of them including an edge (red) in the parallel edge set of the sheet. Note that the relationship between non-hex elements and sheets is important, as non-hex elements interconnect sheets.

\subsection{Sheet Classification}

To identify sheets and their associated edges, we adopt the concepts of \emph{matching} from graph theory \cite{west2001introduction,gibbons1985algorithmic}. Theoretically, in a graph $G$, a matching $M$ is a set of edges that \emph{do not} share any vertices. A vertex in $M$ is \emph{matched} if it is an endpoint of one of the edges in the matching. Otherwise, the vertex is called \emph{unmatched}. 

\begin{figure}
     \centering
     \begin{subfigure}[b]{0.3\columnwidth}
         \centering
         \includegraphics[width=0.9\columnwidth]{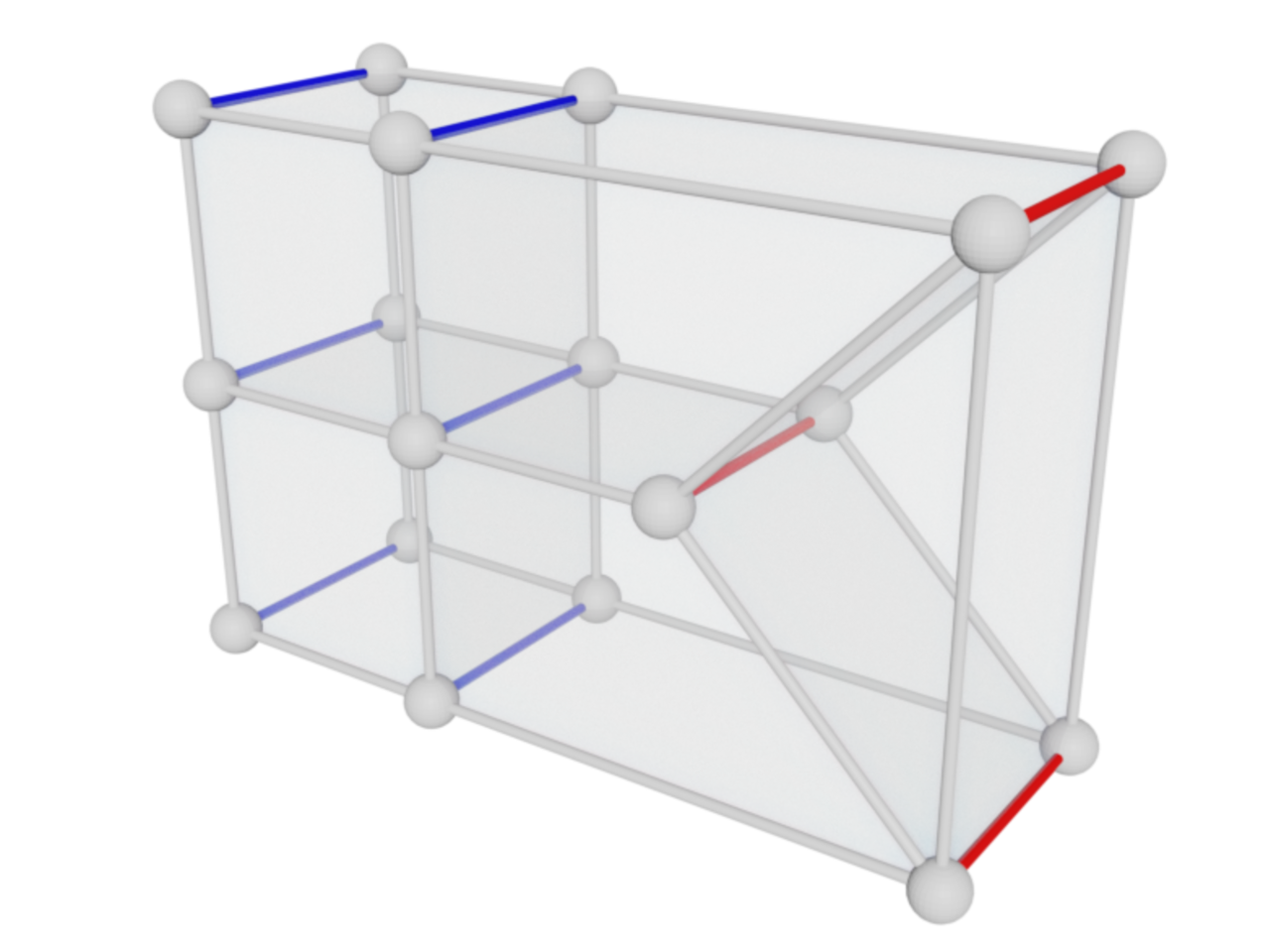}
         \caption{}
         \label{fig:perfect_sheet}
     \end{subfigure}
     \hspace{0.01\textwidth}
     \begin{subfigure}[b]{0.3\columnwidth}
         \centering
         \includegraphics[width=0.9\columnwidth]{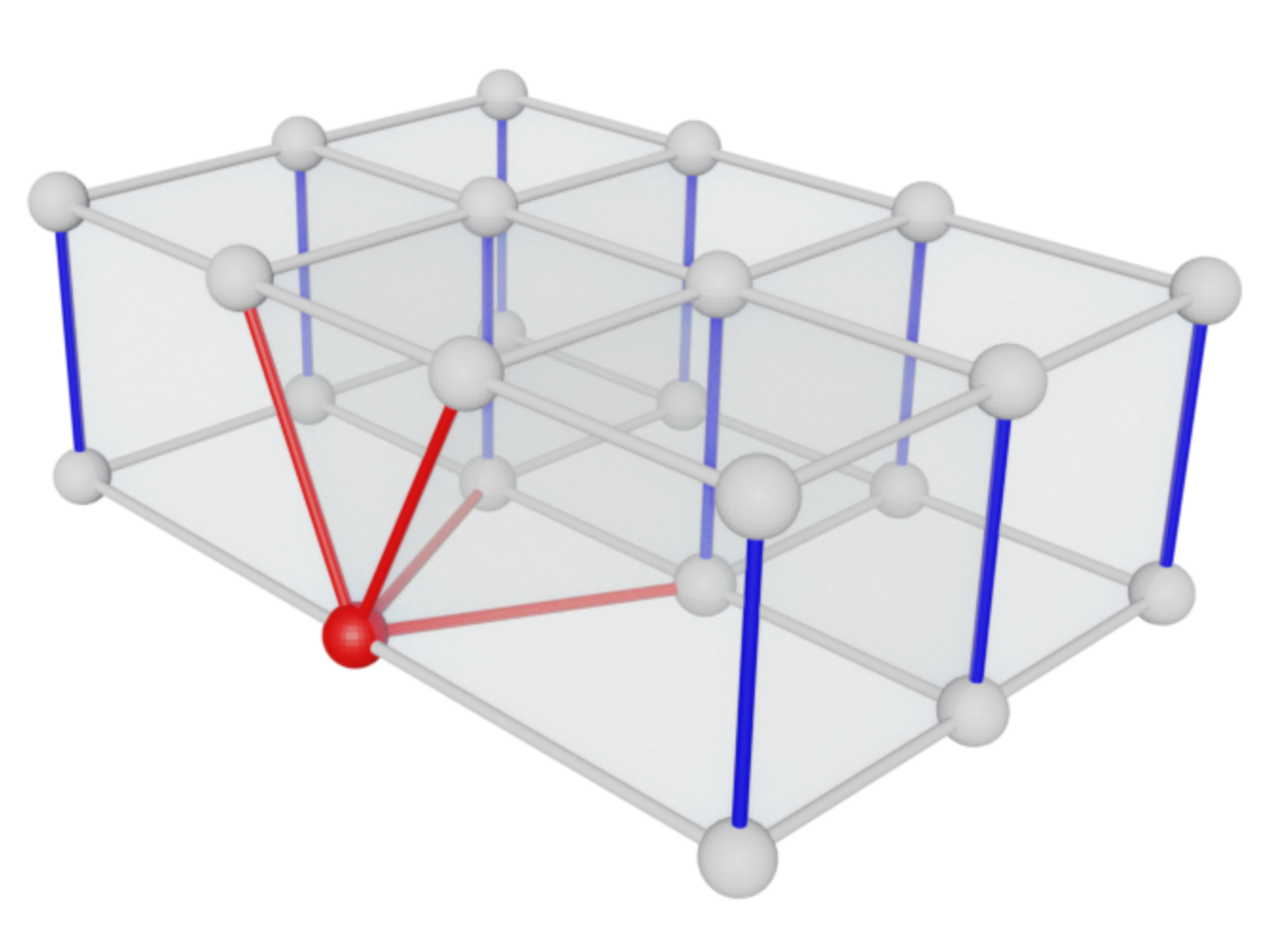}
         \caption{}
         \label{fig:imperfect_sheet}
     \end{subfigure}
     \hspace{0.01\textwidth}
     \begin{subfigure}[b]{0.3\columnwidth}
         \centering
         \includegraphics[width=0.95\columnwidth]{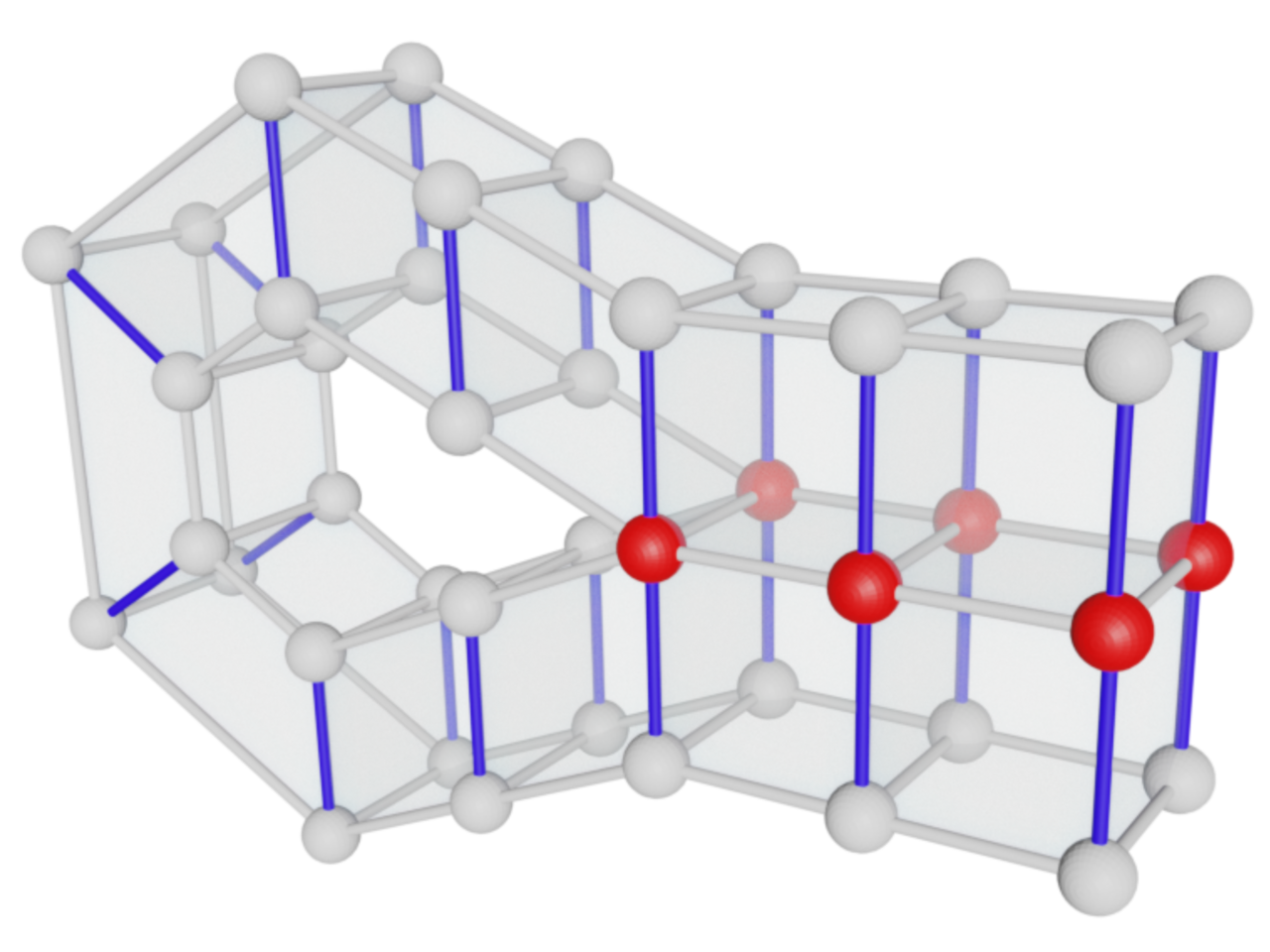}
         \caption{}
         \label{fig:self_parallel}
     \end{subfigure}
     
     \begin{subfigure}[b]{0.3\columnwidth}
         \centering
         \includegraphics[width=0.95\columnwidth]{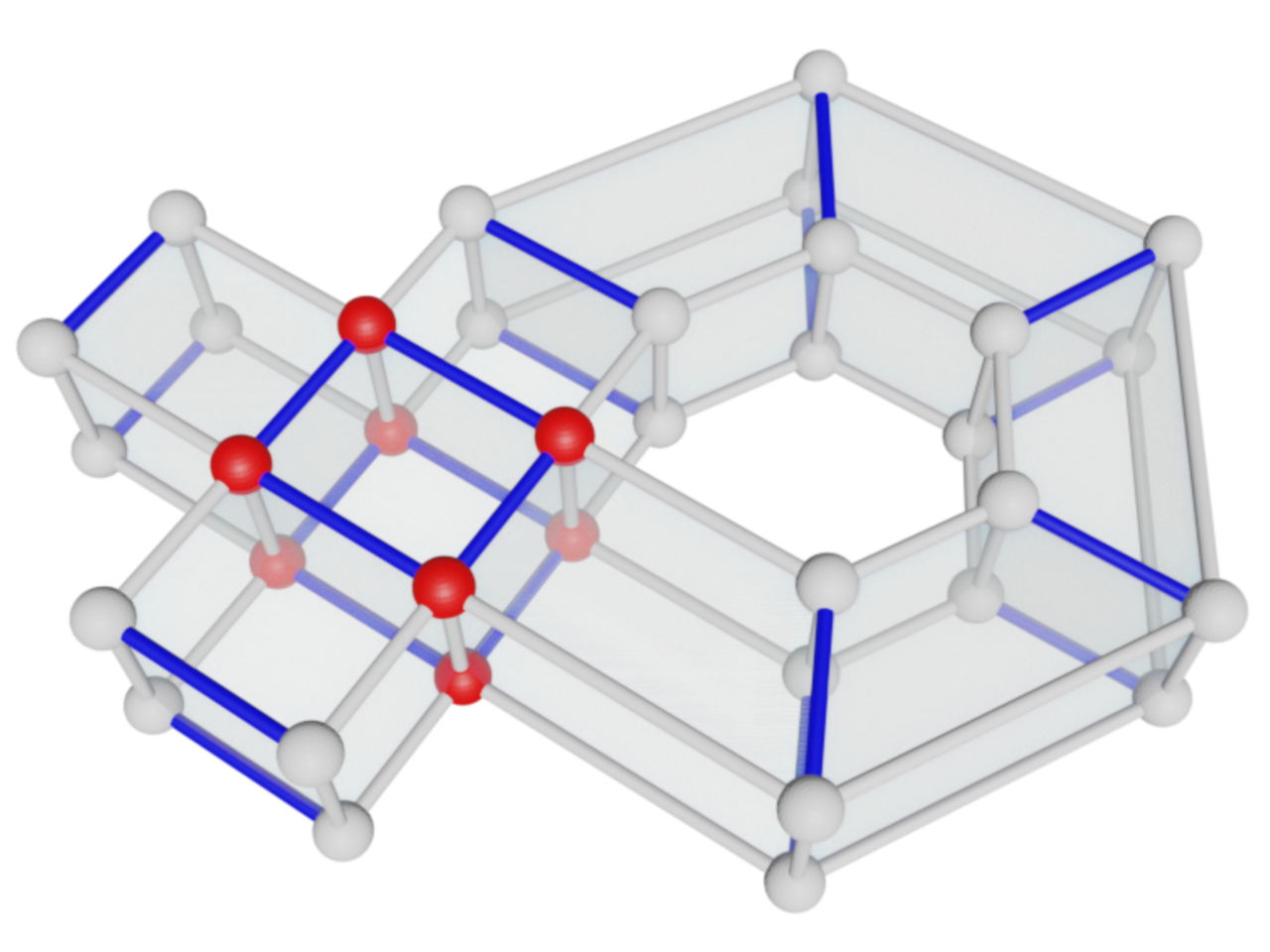}
         \caption{}
         \label{fig:self_intersecting}
     \end{subfigure}
     \hspace{0.01\textwidth}
     \begin{subfigure}[b]{0.3\columnwidth}
         \centering
         \includegraphics[width=0.95\columnwidth]{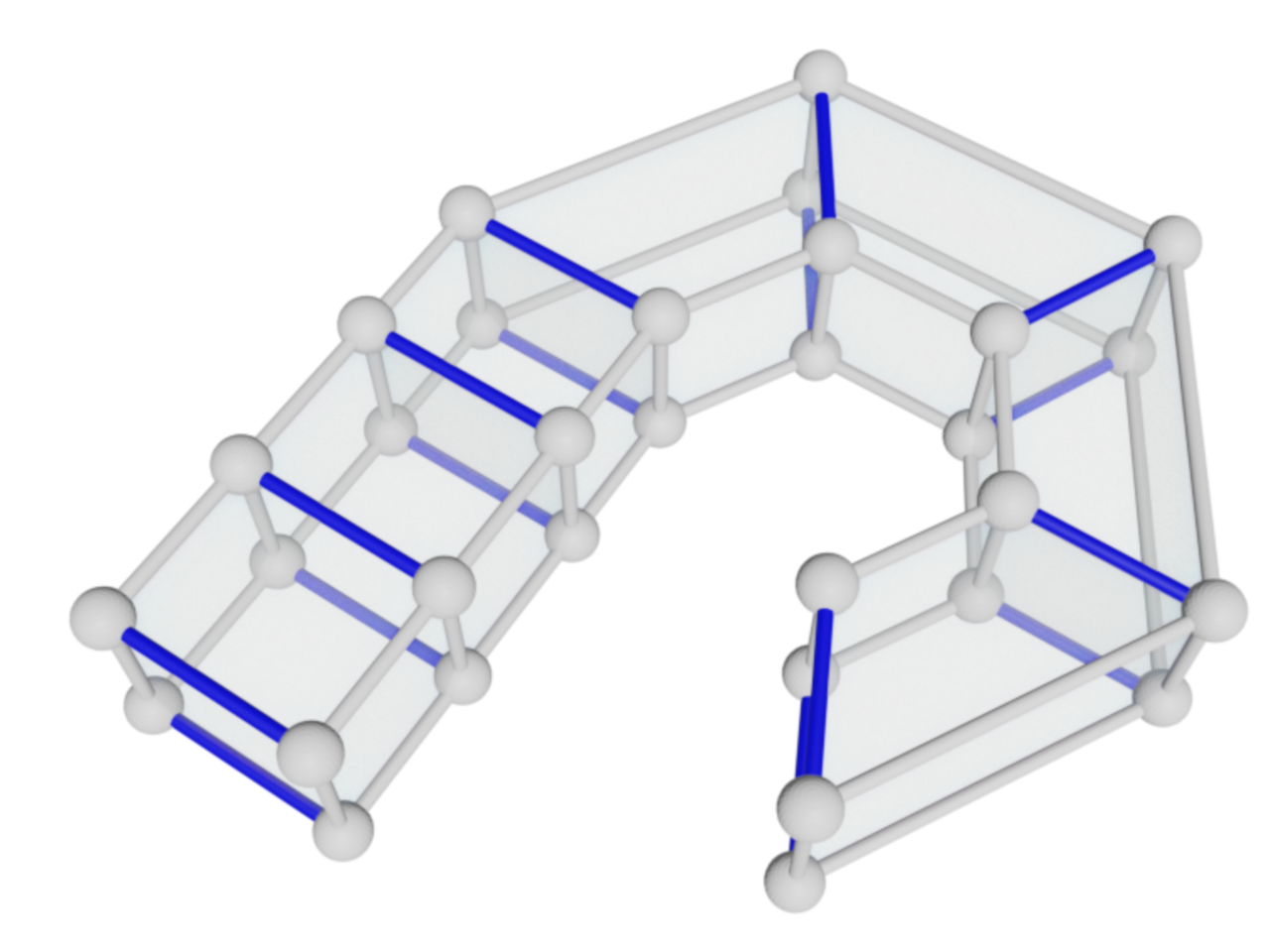}
         \caption{}
         \label{fig:self_intersecting_1}
     \end{subfigure}
     \hspace{0.01\textwidth}
     \begin{subfigure}[b]{0.3\columnwidth}
         \centering
         \includegraphics[width=0.95\columnwidth]{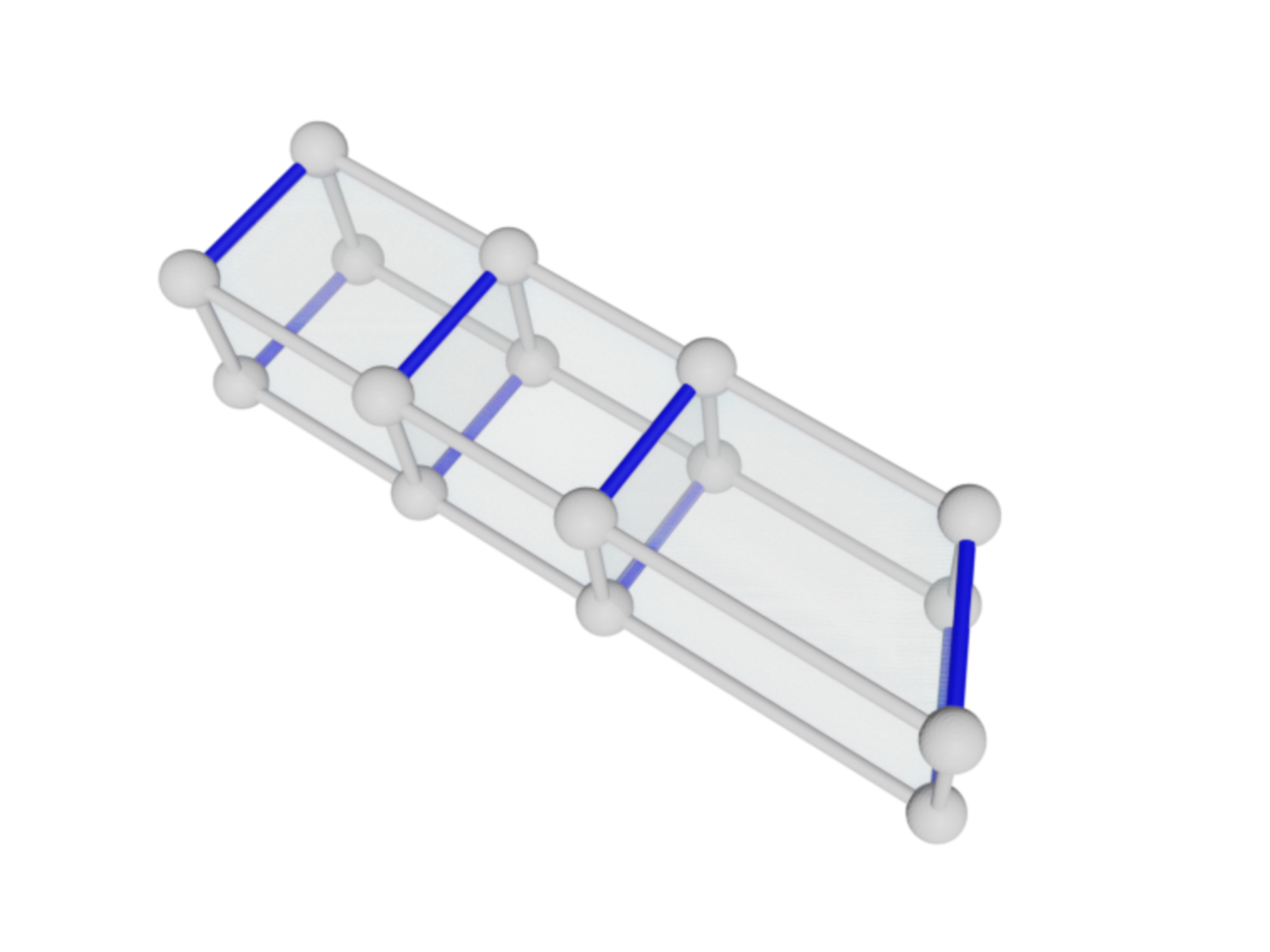}
         \caption{}
         \label{fig:self_intersecting_2}
     \end{subfigure}
     
    \caption{Sheet Types: (a) is a perfect sheet with a non hex neighbor, the red edges means the edge is both parallel edges in the sheet and also part of non hex cell. (b) type-1 imperfect sheet, (c) self-parallel (type-2 imperfect) sheet, (d) self-intersecting (type-3) sheet, (d) could be further split into (e) and (f). Both sub-sheets are perfect sheets. Red vertices here indicate the unmatched vertices.}
    \label{fig:matching_sheet_type}
\end{figure}

With the concept of matching mentioned above, we can now classify sheets extracted from a hybrid base complex. Note that the parallel edges that define a sheet span from one end to the other. 
When all these edges do not share any vertices, they are considered \emph{matched} based on the above matching theory, as illustrated by the colored edges in \autoref{fig:perfect_sheet}. Such sheets are labeled as \emph{perfect sheets}. 
The sheets whose parallel edges cannot be all included in a matching (i.e., some share the same vertices) are referred to as \emph{imperfect sheets}.
For the imperfect sheets, three different configurations can be determined by the neighbor relationship of parallel edges.

\begin{enumerate}
    \item \emph{Type 1 imperfect configuration} shown in \autoref{fig:imperfect_sheet} is caused by some parallel edges sharing vertices due to non-hex elements.
    \item \emph{Type 2 imperfect configuration} shown in \autoref{fig:self_parallel} is the configuration where at least two parallel edges that share a vertex are in two different but adjacent hex cells. The sheets having this configuration are called \emph{self-parallel sheets}.
    \item \emph{Type 3 imperfect configuration} shown in \autoref{fig:self_intersecting} occurs when two parallel edges share the same hex cell and are neighboring to each other (i.e. the sheet returns to the same hex cell). The sheets that include this configuration are called \emph{self-intersecting sheets}.
\end{enumerate}

An imperfect sheet in a hex-dominant mesh may contain one or more of the above imperfect configurations. 
During visual analysis, the edges connected to unmatched vertices in imperfect sheets, \autoref{fig:imperfect_sheet}, require special attention, as they often lead to configurations that complicate the subsequent non-hex element reduction processes.

\subsection{Self-intersecting Sheets Decomposition}
\label{sec:sheetdecomp}

Analyzing the structure of self-intersecting sheets remains a challenge. This is because the orientation of the sheets changes at intersections (\autoref{fig:matching_sheet_type}d), complicating the visual analysis of the internal structural configurations (i.e., hard to trace visually).
To reduce the complexity of self-intersecting sheets, we design a sheet decomposition algorithm. This algorithm breaks down the sheets (esp. type-3 imperfect sheets) into smaller subsheets, as illustrated in \autoref{fig:self_intersecting}, \autoref{fig:self_intersecting_1}, and \autoref{fig:self_intersecting_2}. It ensures that each subsheet does not contain two parallel edges adjacent to each other within a hex cell (i.e., self-intersecting configuration).
To ensure this, the propagation is performed using the breadth-first search strategy. 

Specifically, given a subsheet $G_{SH*}$, its current parallel edge set is marked as $E_{SP*}$. Given an edge $e_i \notin E_{SP*}$, check whether $e_i$ is adjacent to other parallel edges in $E_{SP*}$. If $e_i$ is not adjacent to any edge from $E_{SP*}$ within a hex cell, add $e_i$ into the $E_{SP*}$, and search for the next edge $e_j$ that $e_j \parallel e_i$. After having a complete parallel edge set $E_{SP*}$, the hex cells whose vertices are visited by the edges in $E_{SP*}$ are considered valid subsheet cells. The edges that have no neighboring cell in the sub-sheet will be removed.

By incorporating the decomposition algorithm, these subsheets are either perfect or contain only type-1 or type-2 imperfect configurations, significantly simplifying the analysis process.

\section{Valence-based Singularity Graph}
\label{sec:vsg}

\begin{figure}[!ht]
     \centering
     \begin{subfigure}[b]{0.19\columnwidth}
         \centering
         \includegraphics[width=\columnwidth]{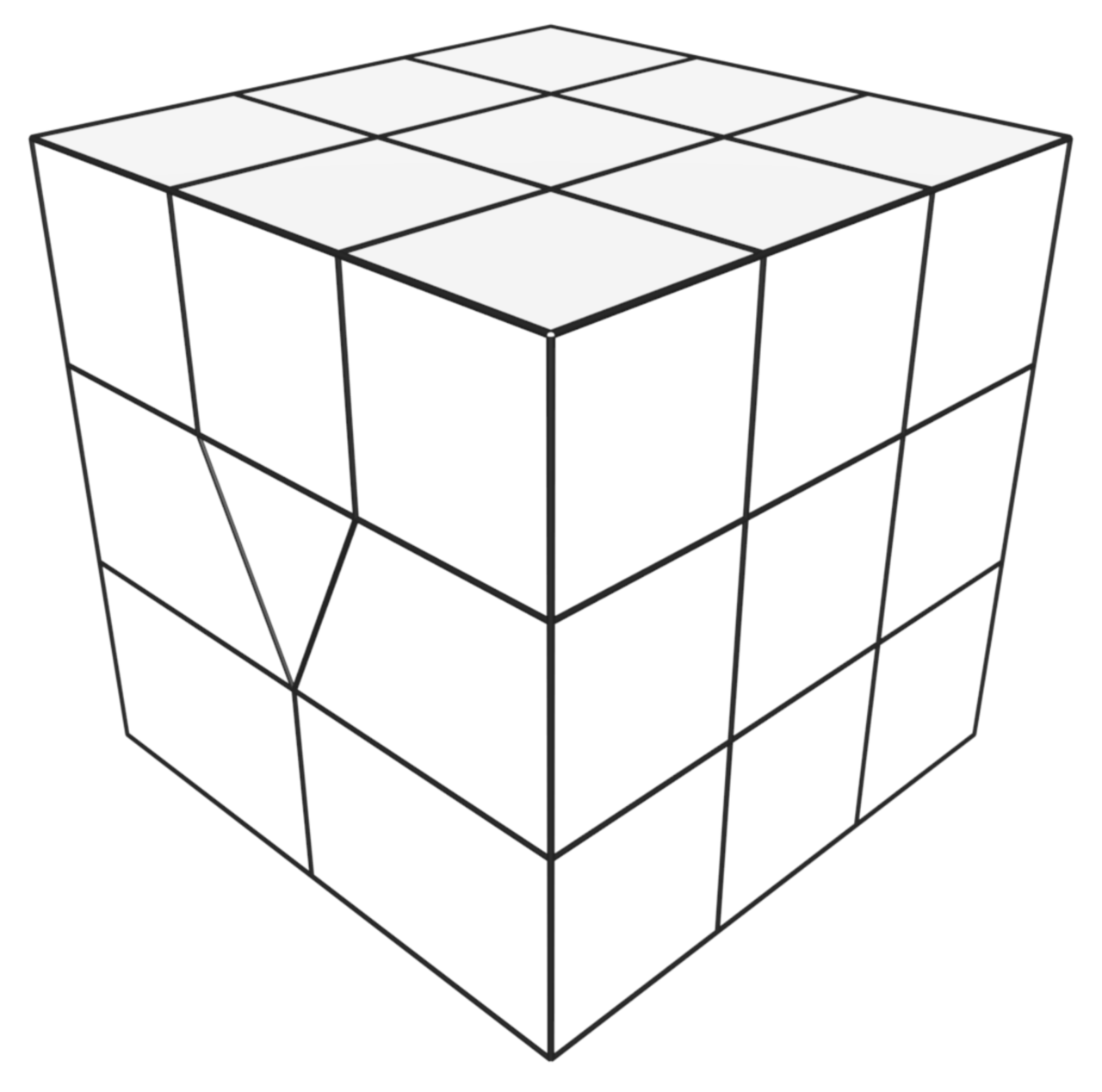}
         \caption{}
         \label{fig:valence-base-sg-surface}
     \end{subfigure}
     \begin{subfigure}[b]{0.19\columnwidth}
         \centering
         \includegraphics[width=\columnwidth]{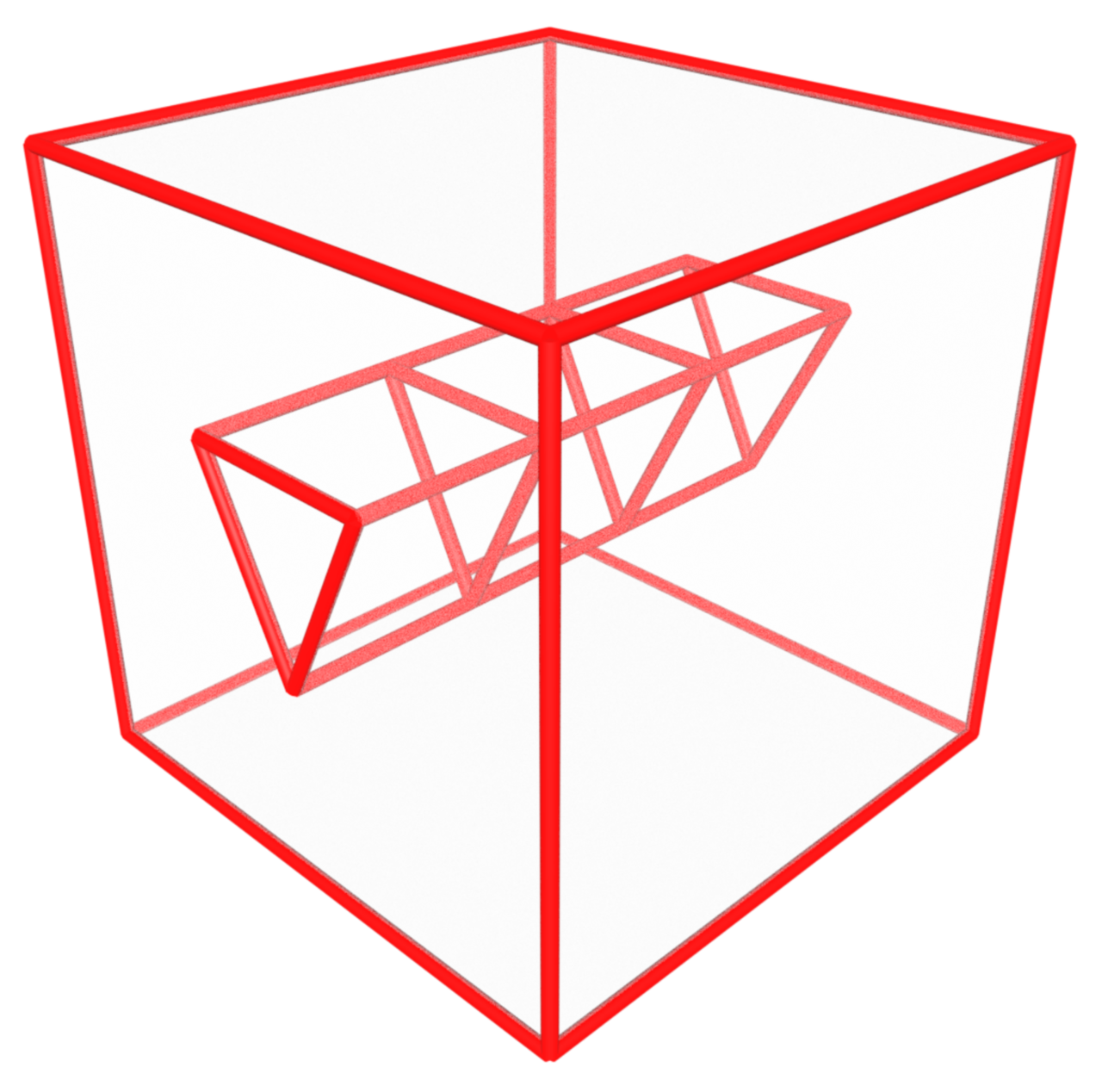}
         \caption{}
         \label{fig:valence-base-sg-hybrid-sg}
     \end{subfigure}
     \begin{subfigure}[b]{0.19\columnwidth}
         \centering
         \includegraphics[width=\columnwidth]{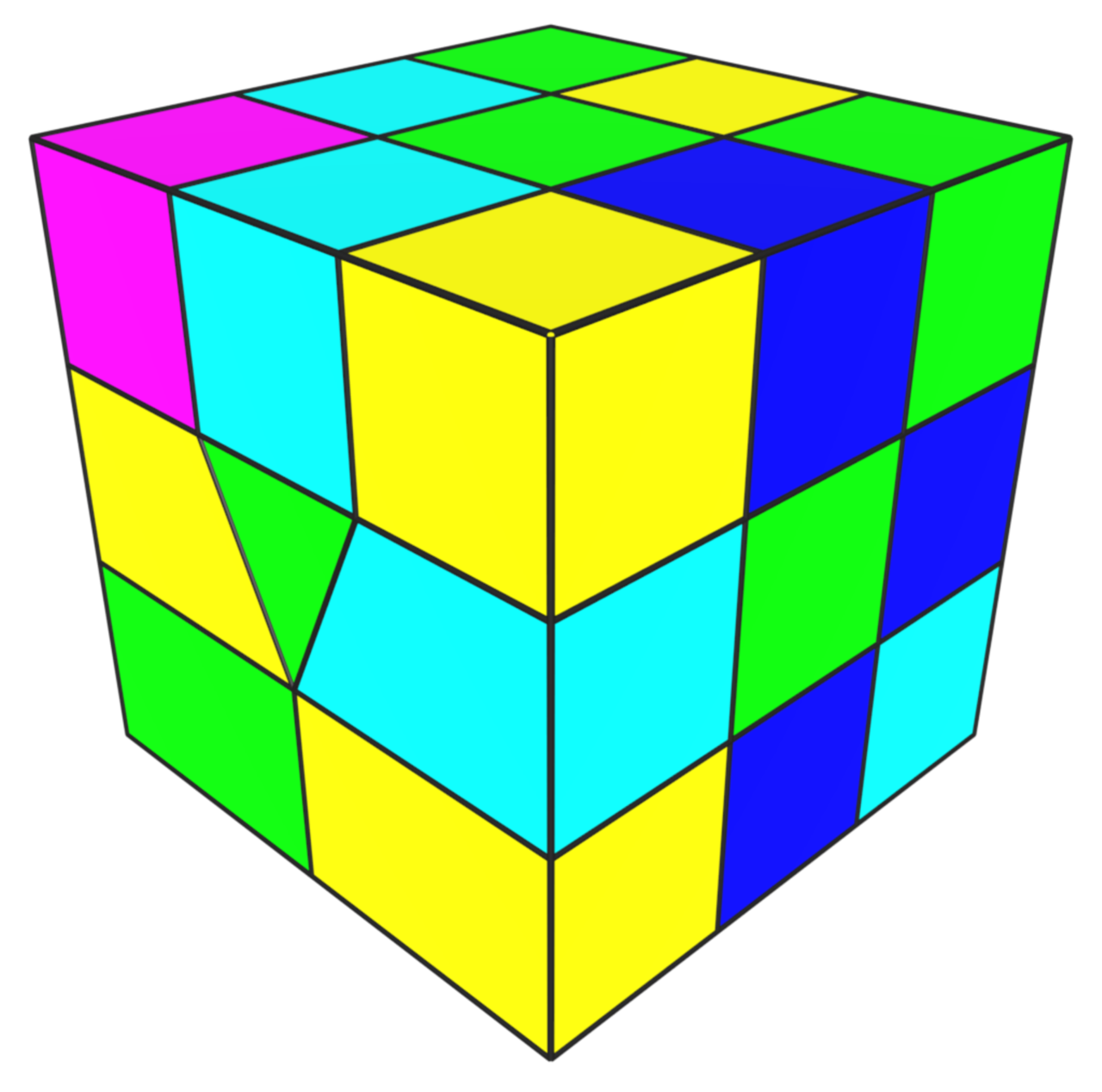}
         \caption{}
         \label{fig:valence-base-sg-hybrid-bc}
     \end{subfigure}
     \begin{subfigure}[b]{0.19\columnwidth}
         \centering
         \includegraphics[width=\columnwidth]{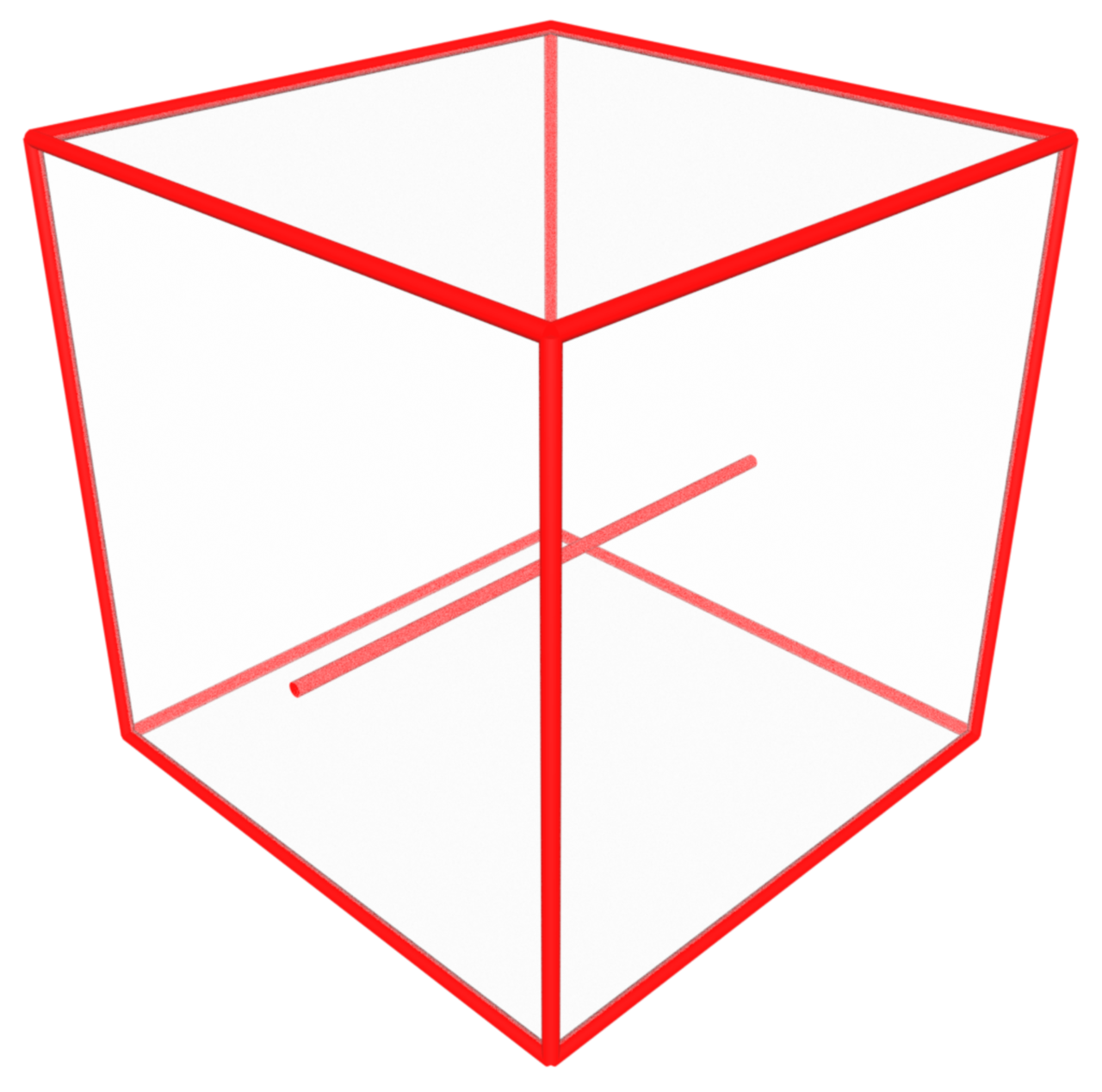}
         \caption{}
         \label{fig:valence-base-sg-see-se}
     \end{subfigure}
     \begin{subfigure}[b]{0.19\columnwidth}
         \centering
         \includegraphics[width=\columnwidth]{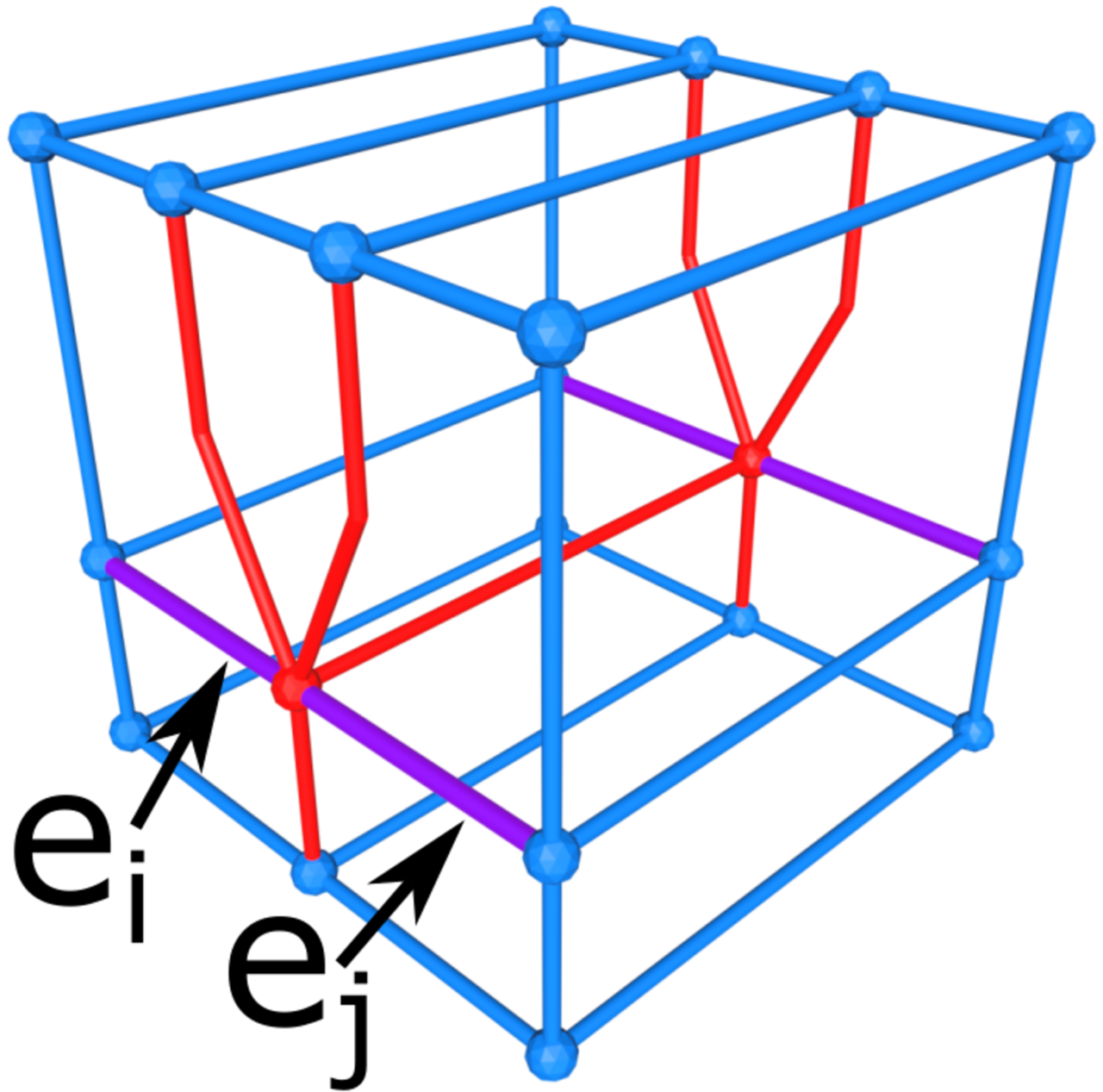}
         \caption{}
         \label{fig:valence-base-sg-complete-sg}
     \end{subfigure}
     \begin{subfigure}[b]{0.19\columnwidth}
         \centering
         \includegraphics[width=\columnwidth]{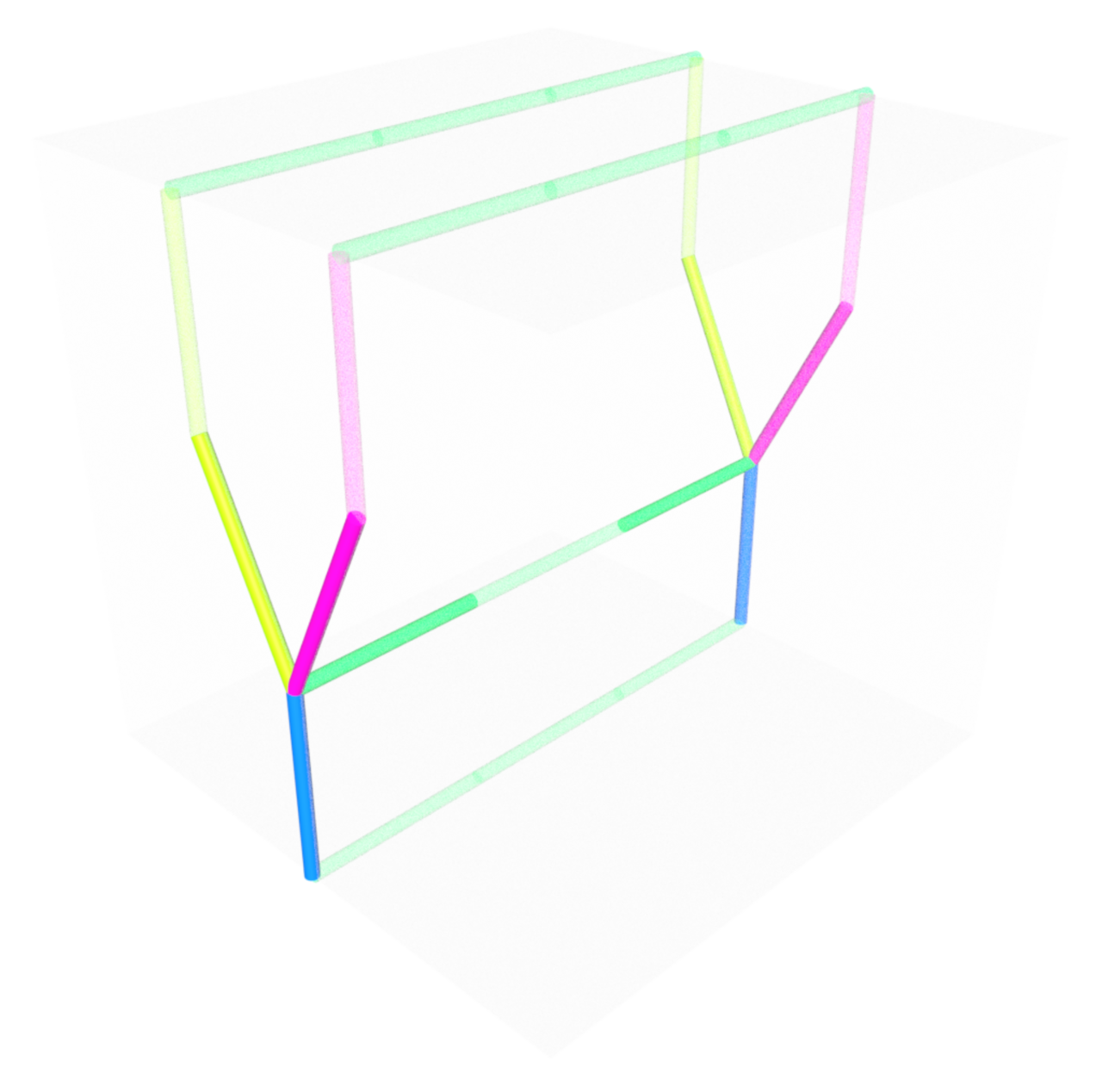}
         \caption{}
         \label{fig:valence-base-sg-cleaned-sg}
     \end{subfigure}
     \begin{subfigure}[b]{0.19\columnwidth}
         \centering
         \includegraphics[width=\columnwidth]{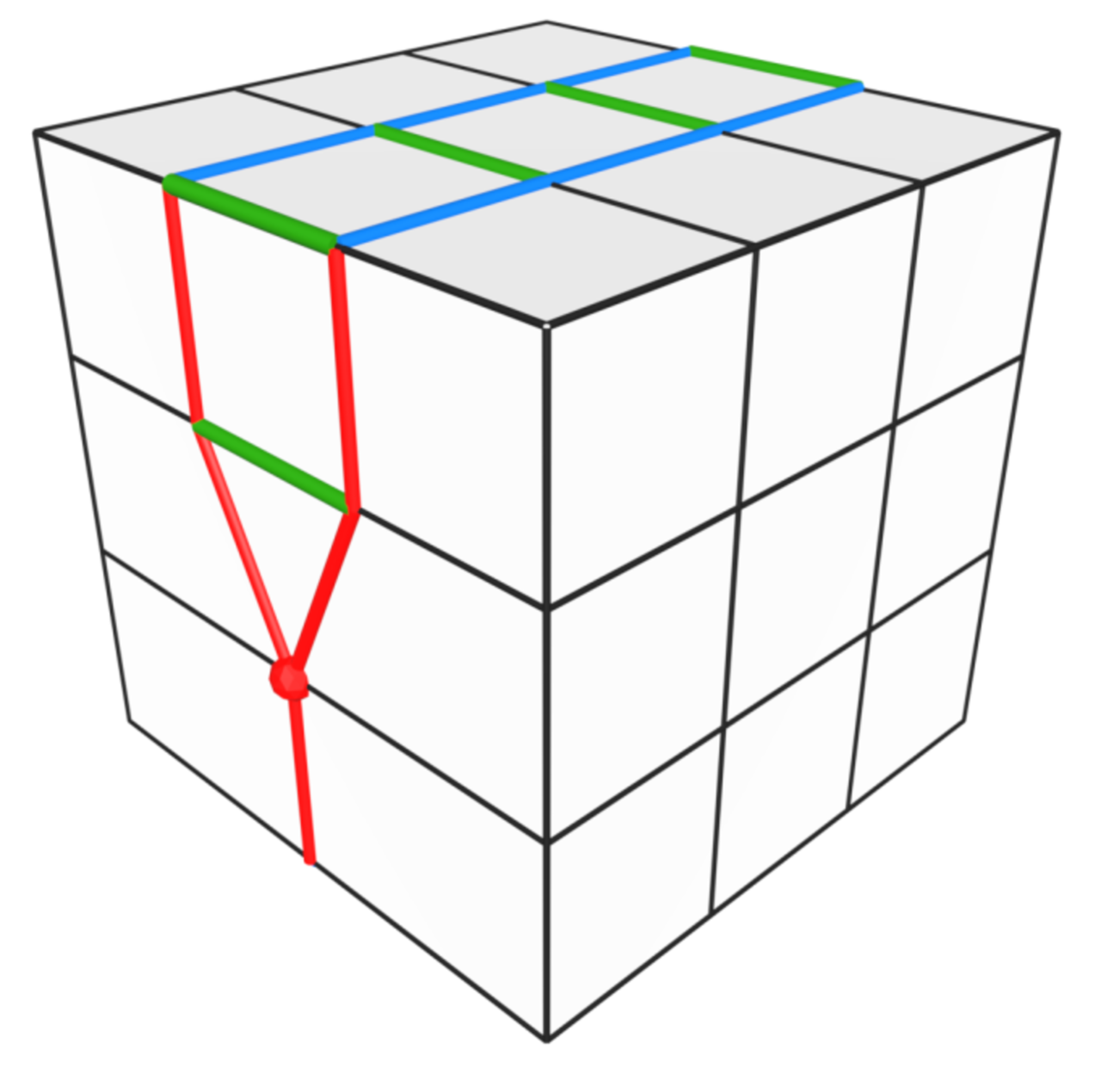}
         \caption{}
         \label{fig:merge_example_1_t1}
     \end{subfigure}
     \begin{subfigure}[b]{0.19\columnwidth}
         \centering
         \includegraphics[width=\columnwidth]{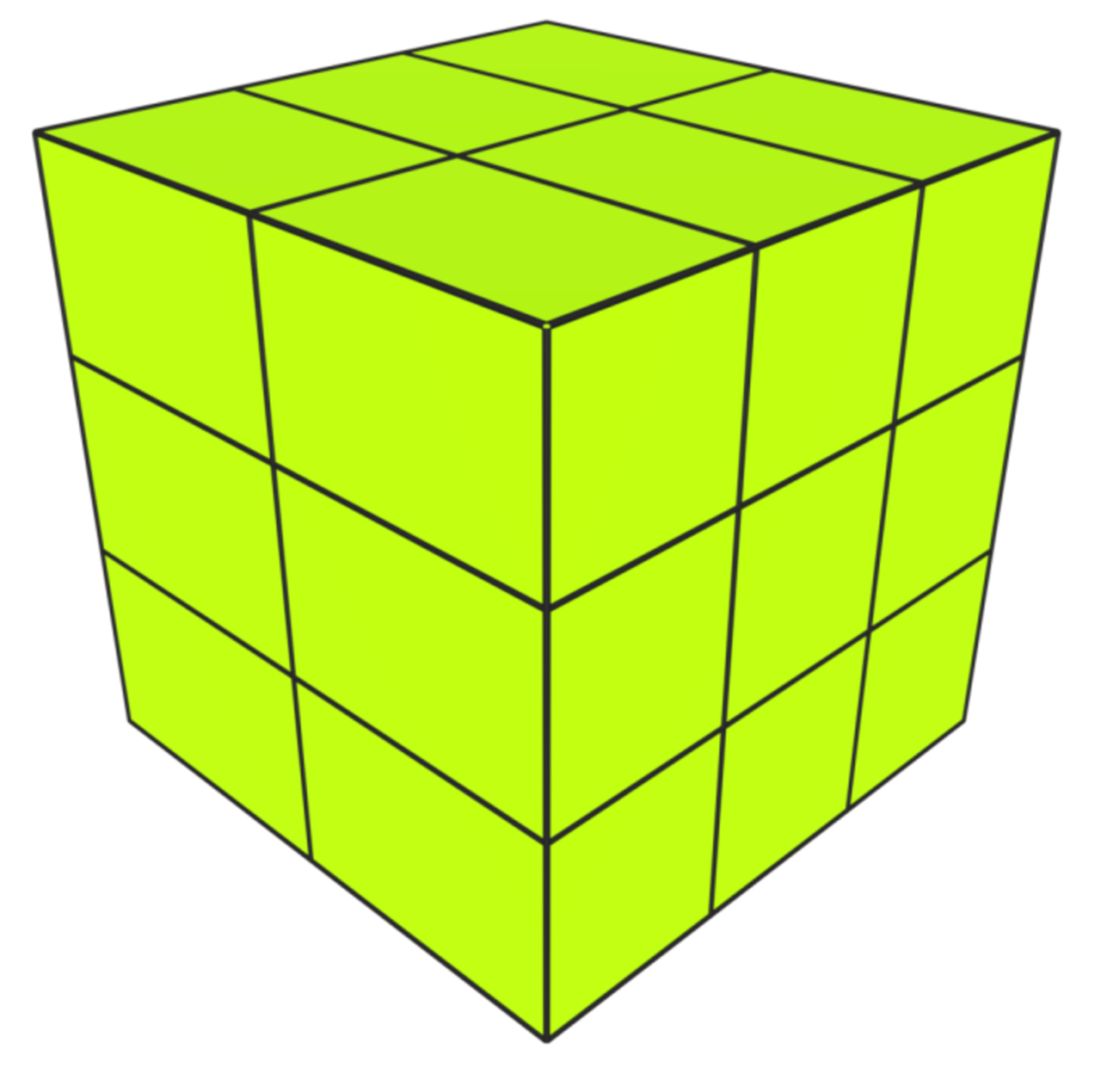}
         \caption{}
         \label{fig:merge_example_1_t2}
     \end{subfigure}
     \begin{subfigure}[b]{0.19\columnwidth}
         \centering
         \includegraphics[width=\columnwidth]{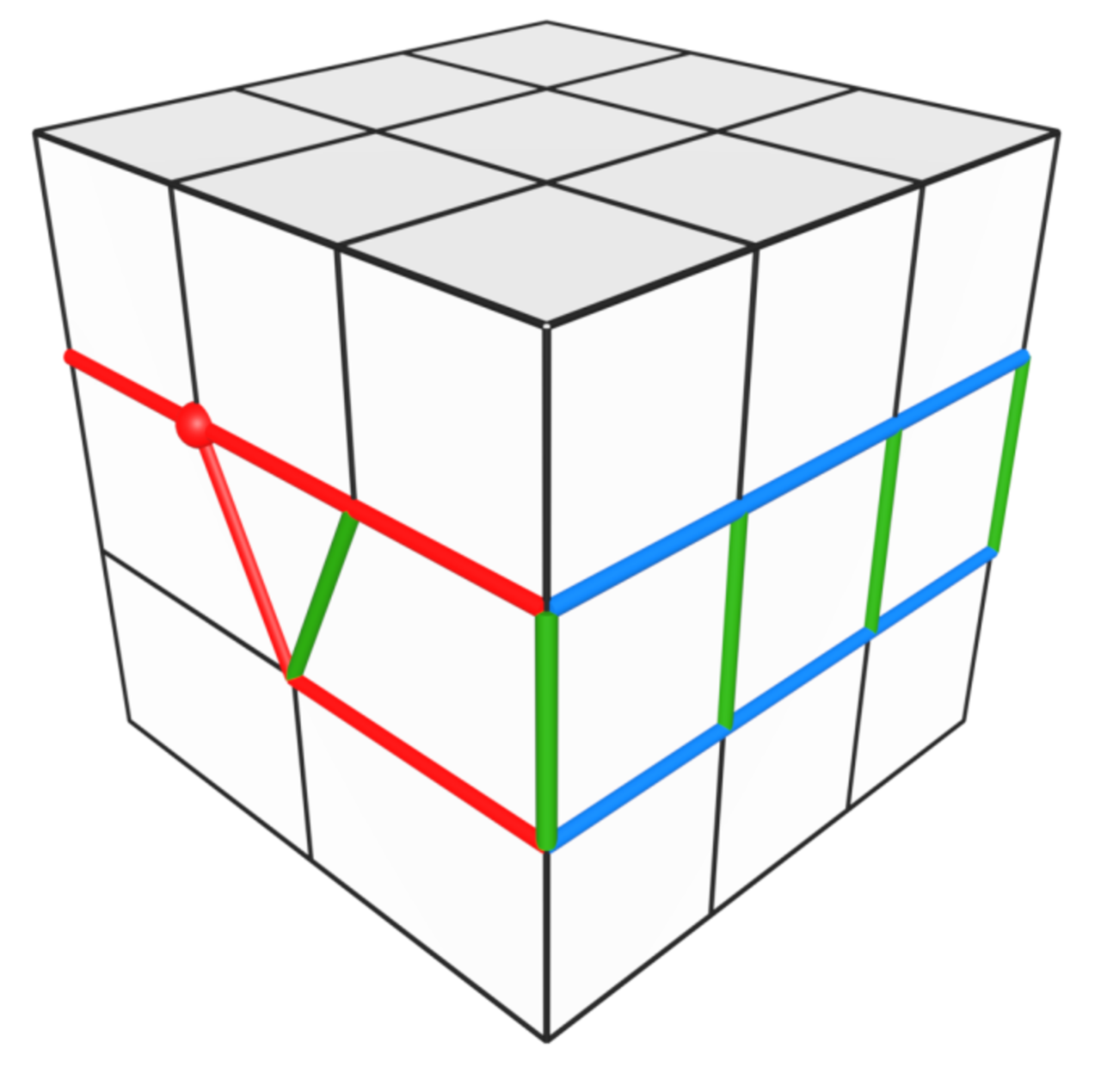}
         \caption{}
         \label{fig:merge_example_2_t1}
     \end{subfigure}
     \begin{subfigure}[b]{0.19\columnwidth}
         \centering
         \includegraphics[width=\columnwidth]{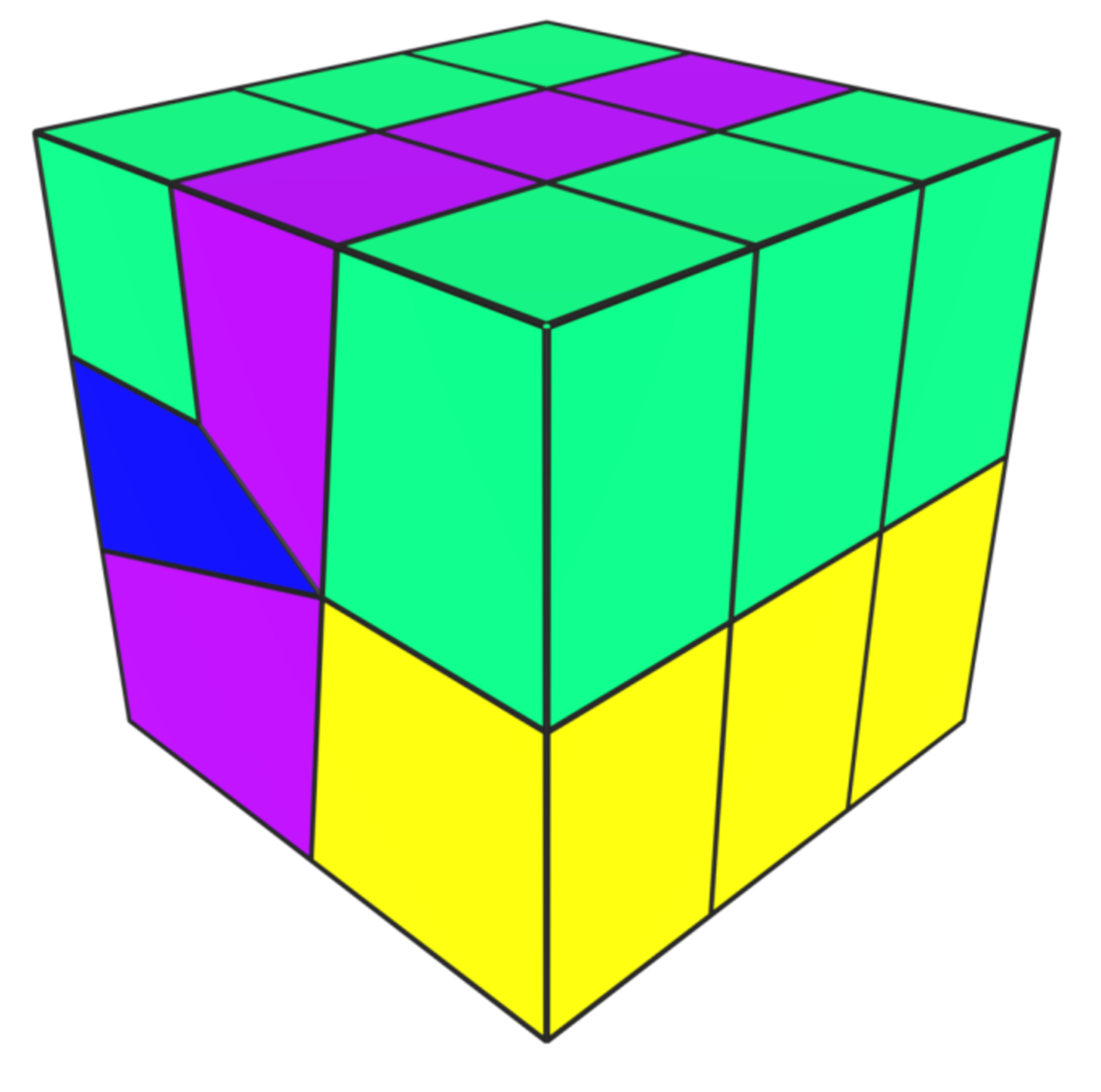}
         \caption{}
         \label{fig:merge_example_2_t2}
     \end{subfigure}
     
    \caption{(a) a simple hex-dominant mesh, (b) its hybrid singularity graph, (c) its hybrid base complex, (d) its $VSG$, (e) the initial complete $VSG$ wireframe, and (f) the (simplified) $VSG$ wireframe. The blue edges (dots) in (e) are the non-important edges (vertices), and the red edges (dots) are important. To retain a connected network for the obtained $VSG$ wireframe, some non-important edges (e.g., the top two middle blue edges) are re-activated.
    (g) shows the sheet suggested by (f) for collapsing. The green edges are the parallel edges of this sheet. After collapsing the ``Y''-shape non-hex configuration (manually), a regular all-hex mesh is resulted with only one hex block (h). (i) shows another sheet for collapsing that can also remove the non-hex configuration. However the resulting all-hex mesh (j) has a complex structure (indicated by multiple colored blocks).   
    }
    \label{fig:valence_based_sg}
\end{figure}
    
Although the proposed hybrid base complex effectively describes the global structure of mesh, the internal structure remains invisible, as illustrated in \autoref{fig:valence-base-sg-hybrid-bc}. In particular, researchers who aim to remove non-hex elements wish to know how certain non-hex configurations (e.g., the triangle-like configuration seen at the boundary of \autoref{fig:valence-base-sg-hybrid-bc}) propagate through the volume to determine the difficulty and impact of their removal. 
One strategy is to look at the wireframe of the hybrid base complex, which may contain too many edges that are not related to irregular and non-hex configurations, complicating this study. 

To address this challenge, we extract a wireframe from a \emph{valence-based singularity graph}, denoted by $VSG$. A $VSG$ consists of all irregular edges in a hex-dominant mesh. A change we made to identify irregular edges, compared to the definition of hybrid singularity structure (or graph) in \autoref{sec:hybridstructure}, is that an edge that is adjacent to a non-hex cell can be part of a singularity if it is irregular (i.e., its valence is not 4 in the interior or not 2 on the boundary).
Now, each singularity includes all connected irregular edges (adjacent to non-hex cell or not) with the same valence. 
The endpoints of each singularity are the singular vertices. All singularities and singular vertices form a valence-based singularity graph ($VSG$) (\autoref{fig:teaser}e). From this $VSG$, we construct a wireframe (\autoref{fig:valence-base-sg-complete-sg}) by tracing along all mesh edges (regular or irregular) from the individual singular vertices until they end at another singularity or at the boundary. We refer to this wireframe as a \emph{$VSG$ wireframe}. This wireframe may still be too complex to analyze, and thus, a further reduction strategy is applied.

\subsection{$VSG$ Wireframe Simplification}
\label{sec:vsgwireframerefinement}

We observe that not all edges in the obtained $VSG$ wireframe contribute to the depiction of the structure configuration, particularly the irregular and non-hex configurations and their propagation. For example, the regular edges adjacent to a corner at the boundary need not depict the non-hex configurations and can be removed. The boundary edges adjacent to the eight corners of the cube mesh in \autoref{fig:valence-base-sg-complete-sg} are examples of these regular edges.
Some regular edges, while connecting to singular vertices, do not contribute to the understanding of the impact of irregular edges and non-hex configurations if they are orthogonal to those configurations. See the set of purple edges near the bottom of the cube in \autoref{fig:valence-base-sg-complete-sg} for some examples. Removing these edges can reduce the visual clutter, leading to a cleaner illustration of the non-hex configuration and its propagation orientation (\autoref{fig:valence-base-sg-cleaned-sg}).
We refer to these edges as \emph{non-important edges}. Next, we propose a strategy to identify these non-important edges for removal.

We first identify the non-important vertices in the $VSG$ wireframe based on the neighboring face and neighboring edge valence.

\subsubsection{non-important vertices} 

A vertex in $VSG$ wireframe is considered non-important if it satisfies one of the following:
\begin{itemize}
    \item it is not a singular vertex
    \item at a boundary corner with all quad neighboring faces 
\end{itemize}
For example, all the blue vertices in \autoref{fig:valence-base-sg-complete-sg} are non-important.

\begin{figure} [h!]
     \centering
     \begin{subfigure}[b]{0.25\columnwidth}
         \centering
         \includegraphics[width=0.99\columnwidth]{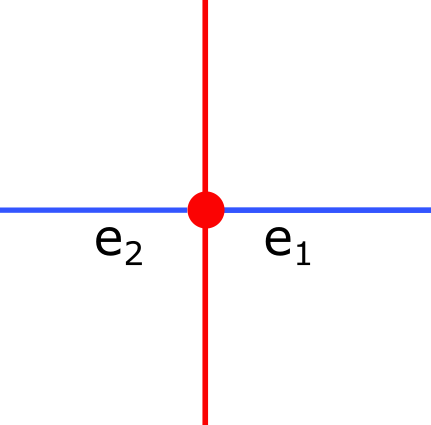}
         \caption{}
         \label{fig:all_non_important_edges}
     \end{subfigure}
     \hspace{0.01\textwidth}
     \begin{subfigure}[b]{0.25\columnwidth}
         \centering
         \includegraphics[width=0.99\columnwidth]{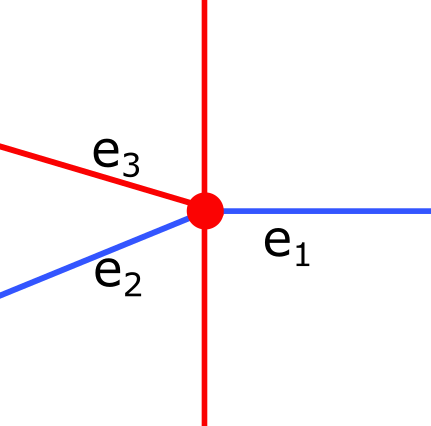}
         \caption{}
         \label{fig:one_important_edges}
     \end{subfigure}
     \hspace{0.01\textwidth}
     \begin{subfigure}[b]{0.25\columnwidth}
         \centering
         \includegraphics[width=0.99\columnwidth]{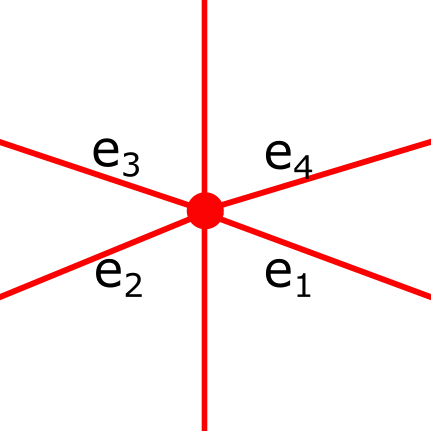}
         \caption{}
         \label{fig:all_important_edges}
     \end{subfigure}
    \caption{
    (a) shows $e_1$ and $e_2$ are non-important edge. (b) shows either $e_2$ or $e_3$ can be paired with $e_1$, but the pair $e_1 \rightarrow_p e_2$ is selected as non-important randomly. (c) In the scenario where all edges surrounding a singular vertex do not have unique paired edges; therefore, all these edges are important.
    }
    \label{fig:non-important-edge-example}
\end{figure}

\subsubsection{non-important edges} 
An edge $e_i$ in a $VSG$ wireframe is determined as non-important using one of the following criteria.

First, if both endpoints (vertices) of $e_i$ are non-important, then $e_i$ is marked as non-important.

Second, given two connected regular edges $e_i$ and $e_j$, if $e_i$ is \emph{paired} with only $e_j$, denoted by $e_i\rightarrow_p e_j$, and $e_j$ is also paired with only $e_i$ (i.e., $e_j\rightarrow_p e_i$), then both $e_i$ and $e_j$ are non-important. Two regular edges adjacent to the same singular vertex can be \emph{paired} if they are from different cells and are adjacent to only quad faces. According to this, $e_1$ and $e_2$ are both non-important in \autoref{fig:all_non_important_edges}.

Third, given a set of regular edges around a singular vertex, if one of these edges, say $e_i$, can be paired with more than one other edge, i.e., a one-to-many pairing, and all the other edges can only be paired with $e_i$, then $e_i$ is non-important. We also randomly mark one of the edges that can be paired with $e_i$ as non-important.  \autoref{fig:one_important_edges} illustrates such a scenario. Specifically, $e_1$ can be paired with both $e_2$ and $e_3$; but $e_2$ can only be paired with $e_1$ and $e_3$ can only be paired with $e_1$, because $e_2$ and $e_3$ are in the same cell. In this case, $e_1$ is non-important, and we randomly choose $e_2$ as another non-important edge.

In the case of \autoref{fig:all_important_edges}, edges $e_1$, $e_2$, $e_3$, and $e_4$ are regular and surrounded by only quad faces. Since all of them can be paired with multiple other edges, none are marked as non-important.

Based on the above criteria, all blue edges in \autoref{fig:valence-base-sg-complete-sg} are non-important, as their endpoints are all non-important (blue). The (purple) edge $e_i$ and $e_j$ are also non-important based on the second criterion above.

\begin{figure} [t]
     \centering
     \begin{subfigure}[b]{0.3\columnwidth}
         \centering
         \includegraphics[width=0.99\columnwidth]{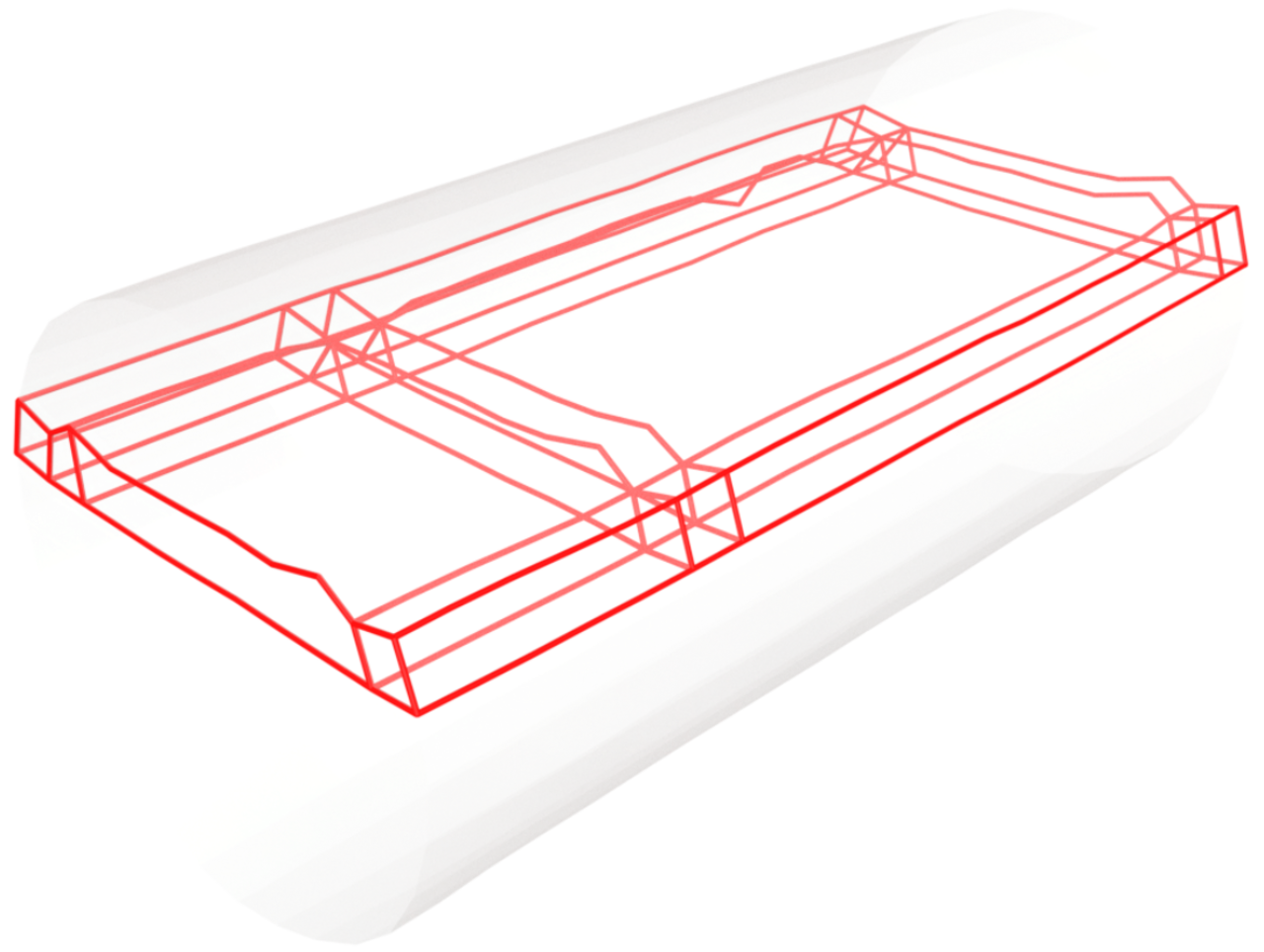}
         \caption{}
         \label{fig:edge_activation_imperfect_sheet}
     \end{subfigure}
     \hspace{0.01\textwidth}
     \begin{subfigure}[b]{0.3\columnwidth}
         \centering
         \includegraphics[width=0.99\columnwidth]{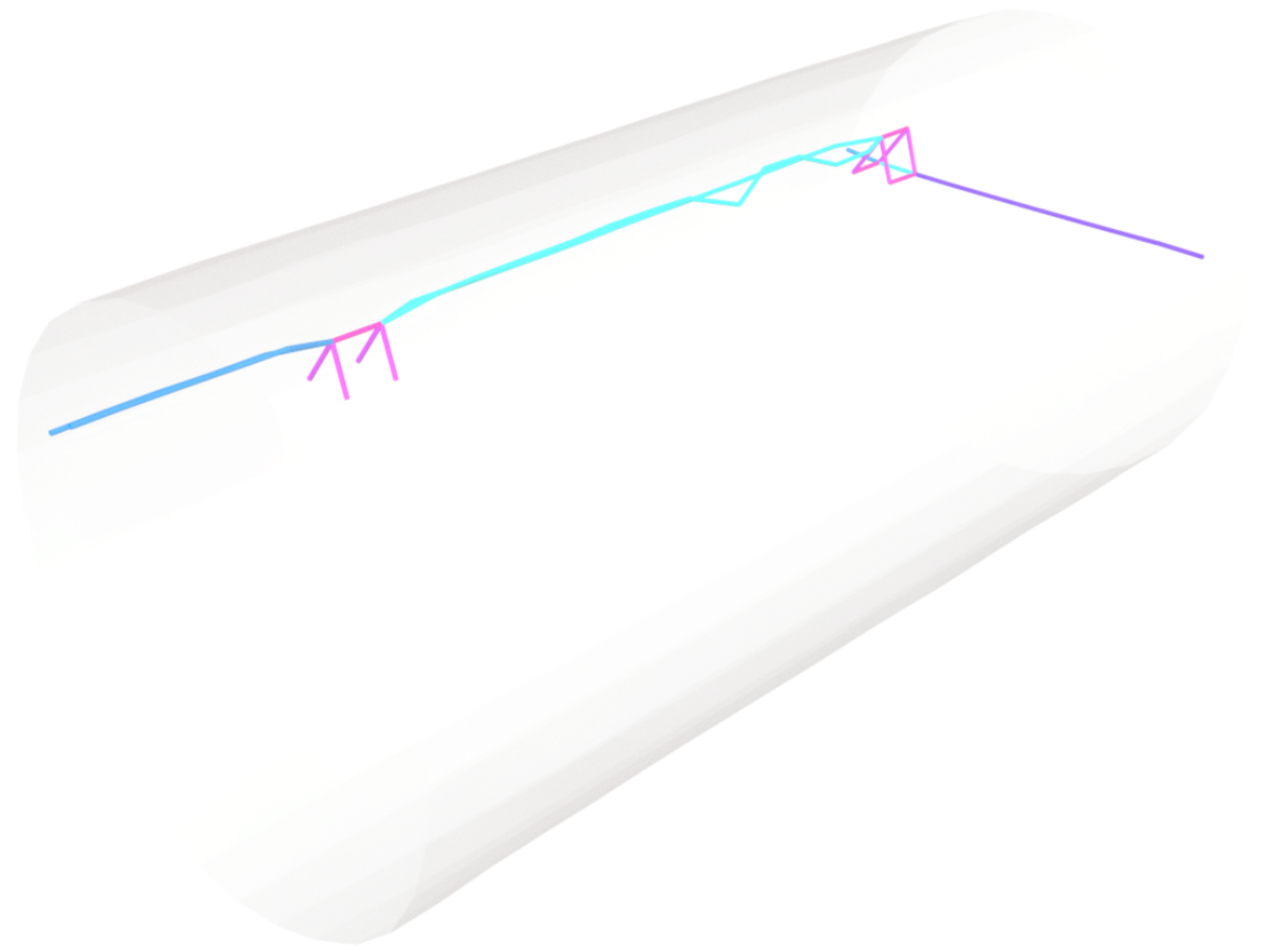}
         \caption{}
         \label{fig:edge_activation_only_singular_edges}
     \end{subfigure}
     \hspace{0.01\textwidth}
     \begin{subfigure}[b]{0.3\columnwidth}
         \centering
         \includegraphics[width=0.99\columnwidth]{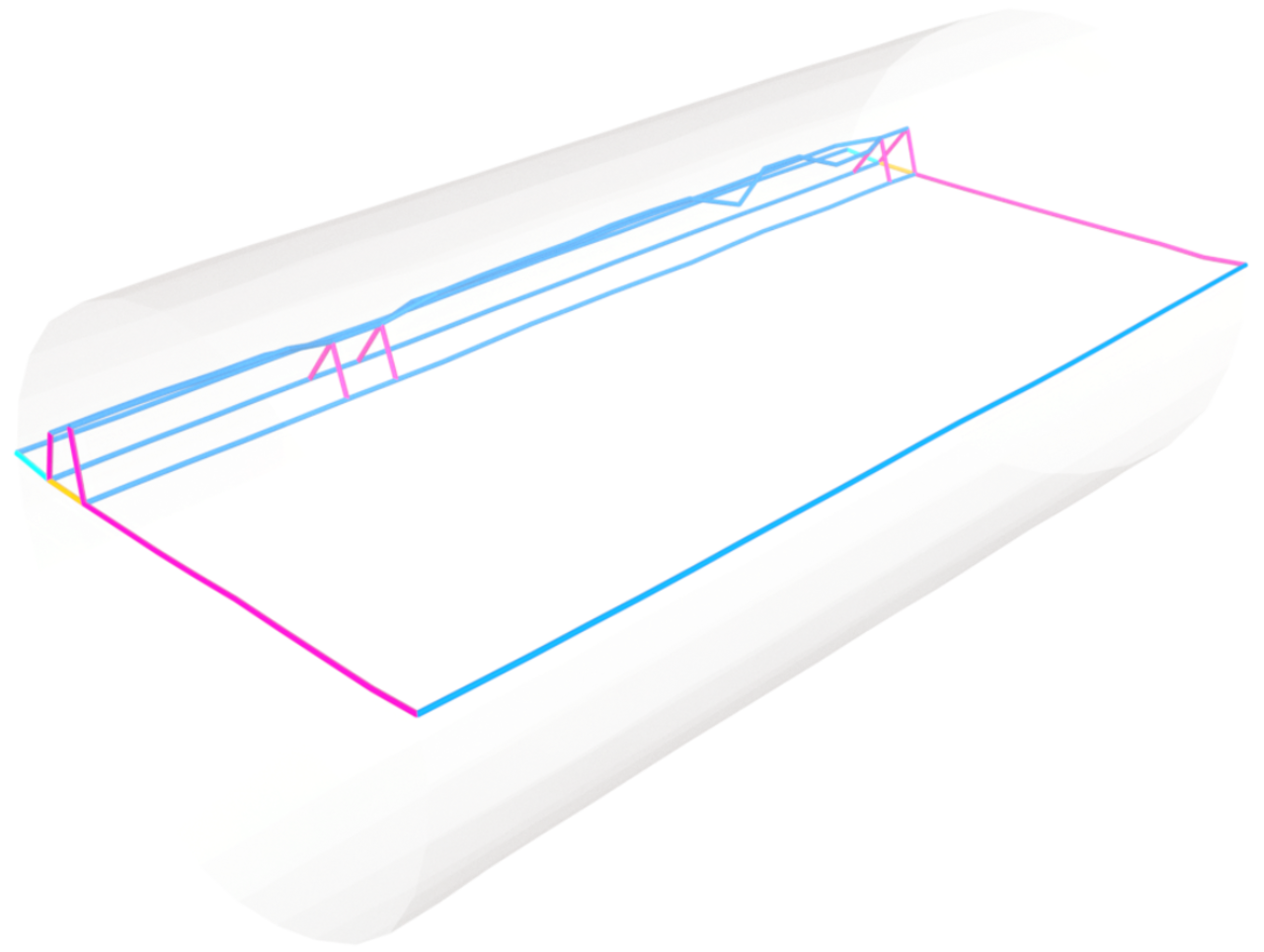}
         \caption{}
         \label{fig:edge_activation_complete}
     \end{subfigure}
    \caption{Given the initial $VSG$ wireframe of a sheet (a), non-important edges are hidden to reduce visual clutter (b). Reactivating some non-important edges is needed to obtain a complete network for the simplified $VSG$ wireframe (c).
    }
    \label{fig:singular_edges_reactivation}
\end{figure}

After detecting non-important elements, they are hidden from the visualization. However, hiding these edges results in a disconnected structure (\autoref{fig:edge_activation_only_singular_edges}), providing incomplete information about its global configuration. To address this, we re-activate some of those edges. In particular, if a non-important edge from the initial $VSG$ wireframe is adjacent to a singular edge, it will be shown in the visualization (\autoref{fig:edge_activation_complete}).

Following the removal and reactivation of non-important edges, we obtain a cleaner and simplified $VSG$ wireframe. This graph effectively highlights the irregular configurations with less clutter, as illustrated in \autoref{fig:valence-base-sg-cleaned-sg}. The orientation of some non-hex configurations in $VSG$ wireframe also suggests a possible ideal direction of the sheet for collapsing to remove these configurations (see \autoref{fig:valence_based_sg}g--j). 
As the rest of the paper will show the simplified $VSG$ wireframes, we will refer to them as $VSG$ wireframes for simplicity.

\section{Visualize Hybrid Base Complex and Its Substructures}

The above sections describe the extraction of various geometric descriptors for the study of the structure of hex-dominant meshes. In this section, we describe our strategies to visualize them.

\begin{figure}[!t]
     \centering
     \begin{subfigure}[b]{0.3\columnwidth}
         \centering
         \includegraphics[width=\columnwidth]{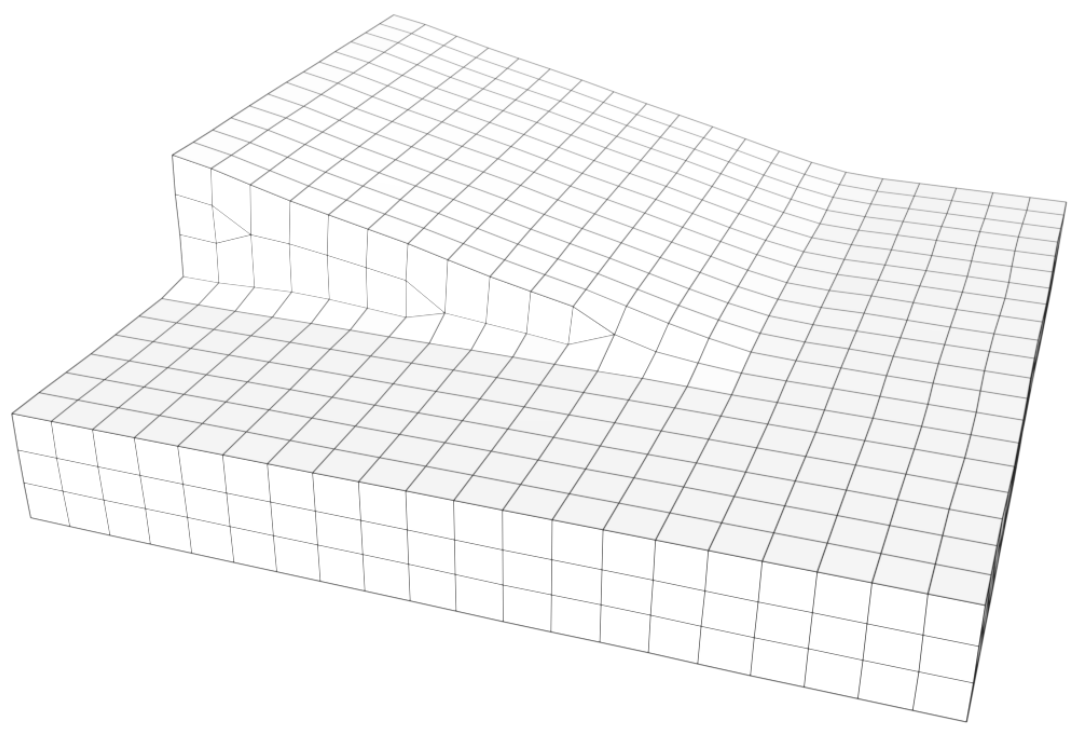}
         \caption{Direct Visualization}
         \label{fig:at_most_jumpramp_surface}
     \end{subfigure}
     \hspace{0.01\textwidth}
     \begin{subfigure}[b]{0.3\columnwidth}
         \centering
         \includegraphics[width=\columnwidth]{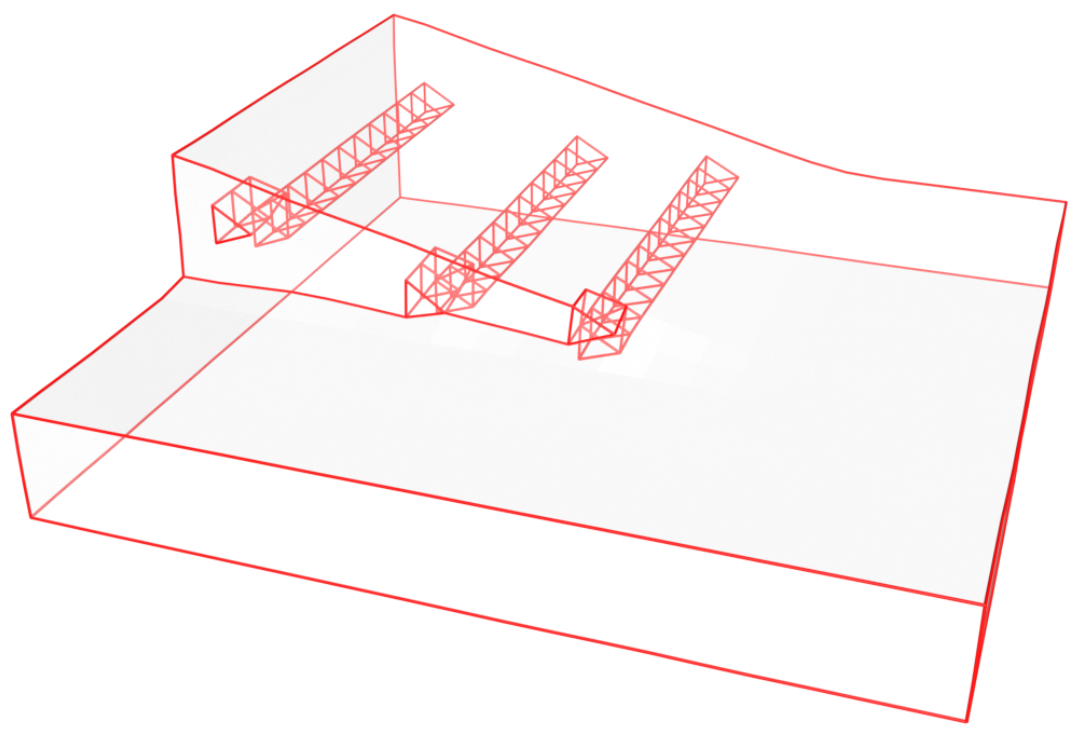}
         \caption{Hybrid Singularity Graph}
         \label{fig:at_most_jumpramp_hsg}
     \end{subfigure}
     \hspace{0.01\textwidth}
     \begin{subfigure}[b]{0.3\columnwidth}
         \centering
         \includegraphics[width=\columnwidth]{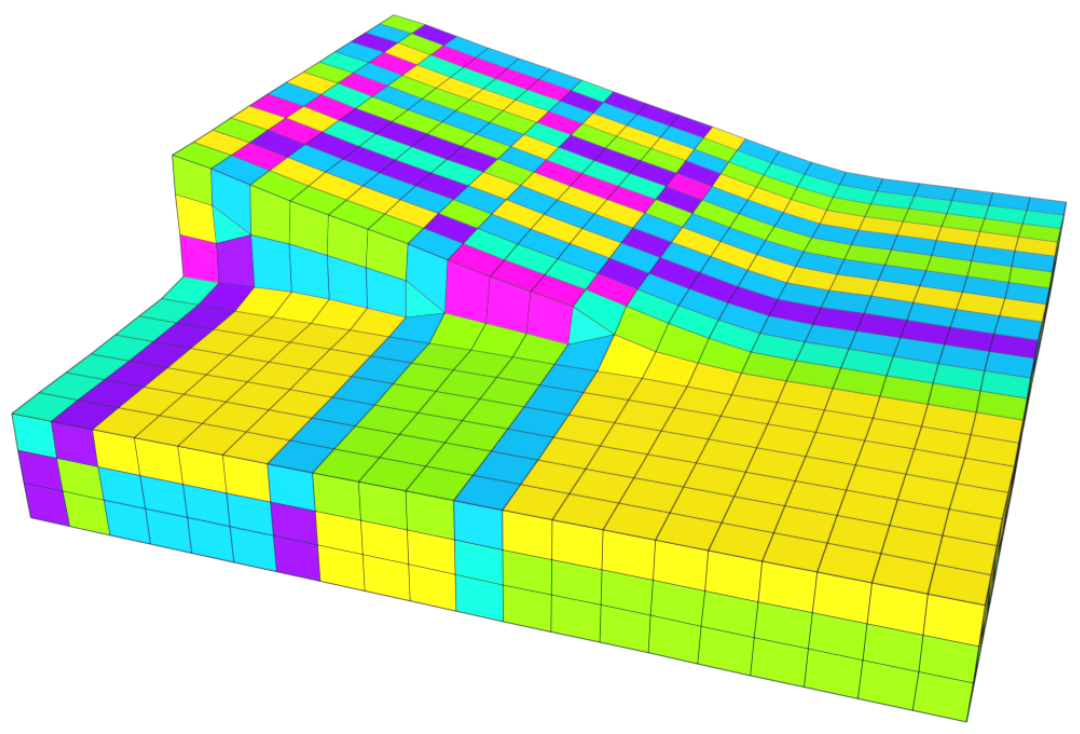}
         \caption{Hybrid Base Complex}
         \label{fig:at_most_jumpramp_hbc}
     \end{subfigure}
    
     \begin{subfigure}[b]{0.3\columnwidth}
         \centering
         \includegraphics[width=\columnwidth]{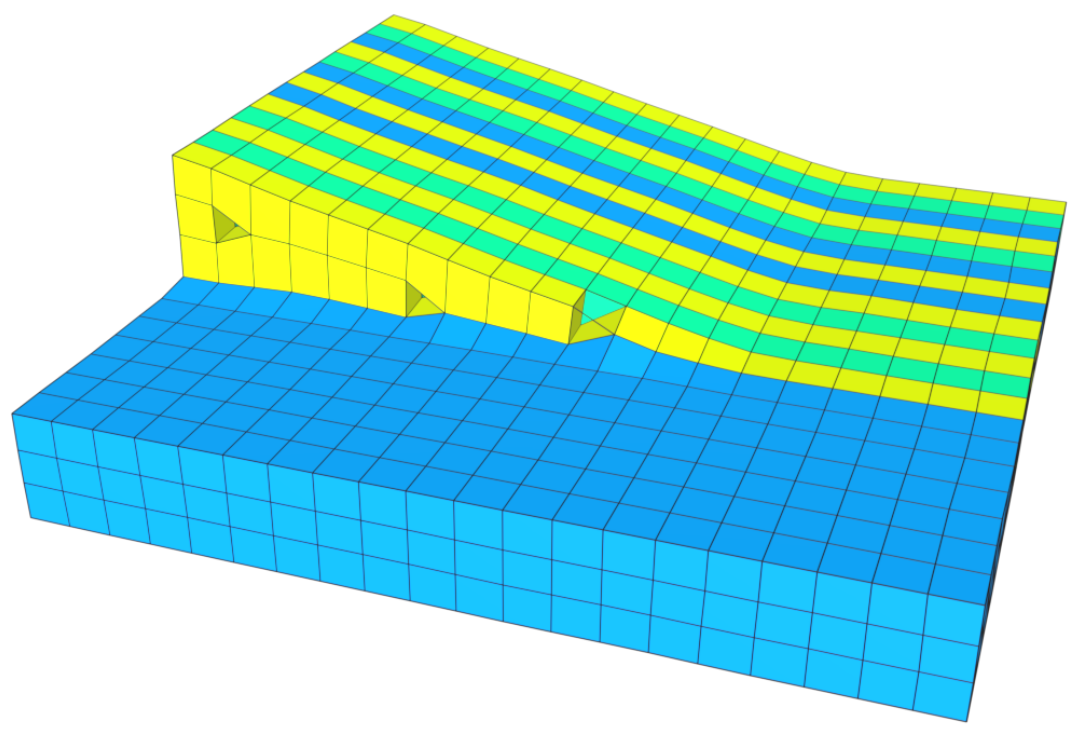}
         \caption{Selected Sheets}
         \label{fig:at_most_jumpramp_sheets}
     \end{subfigure}
     \hspace{0.01\textwidth}
     \begin{subfigure}[b]{0.3\columnwidth}
         \centering
         \includegraphics[width=\columnwidth]{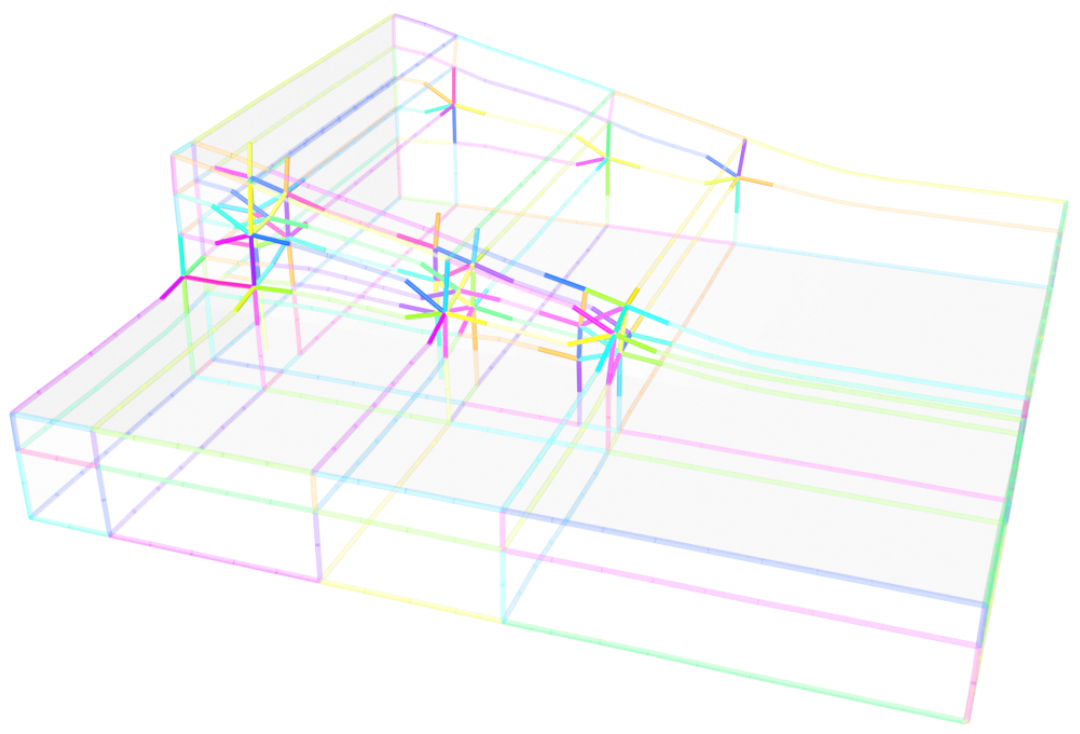}
         \caption{VSG Wireframe}
         \label{fig:at_most_jumpramp_VSG}
     \end{subfigure}
     \hspace{0.01\textwidth}
     \begin{subfigure}[b]{0.3\columnwidth}
         \centering
         \includegraphics[width=\columnwidth]{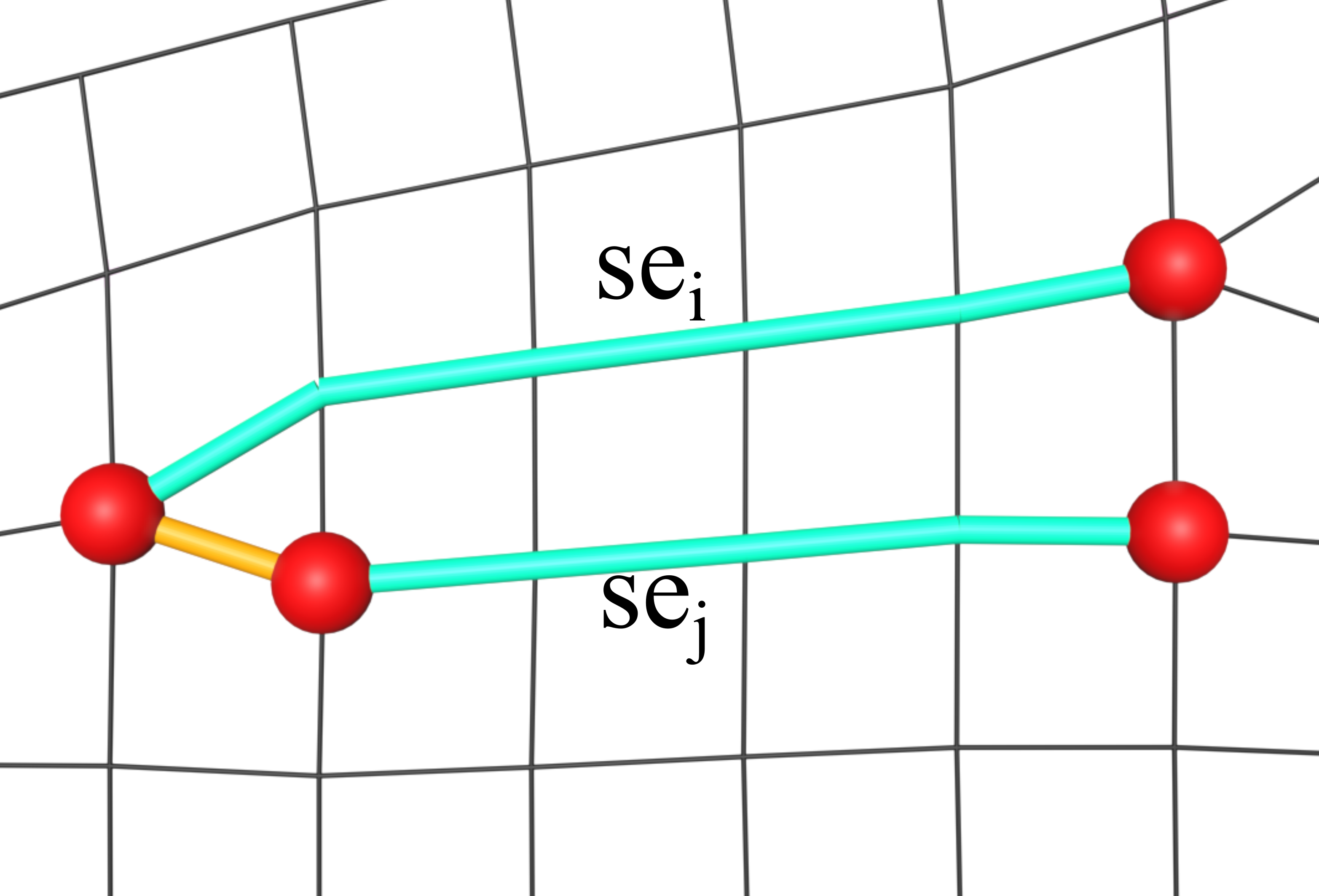}
         \caption{Partial Parallel Singularities}
         \label{fig:at_most_jumpramp_partial_parallel_singularities}
     \end{subfigure}
    \caption{(a) direct visualization of a hex-dominant mesh, (b) its hybrid singularity graph visualized as line graphs, (c) its hybrid base complex visualized using colored blocks, (d) a set of extracted sheets visualized as large colored blocks, (e) its (simplified) $VSG$ wireframe visualized as a 3D network with colors and opacity. (f) illustrates a scenario of partial parallel singularities to determine the colors of edges in the $VSG$ wireframe.
    }
    \label{fig:at_most_jumpramp_example}
\end{figure}

\subsection{Visualize Hybrid Singularity Graph and Base Complex}

A hybrid singularity graph consists of 1D edges and 0D vertices. We visualize them as colored lines. It provides an initial description of the complexity of the structure, as illustrated in \autoref{fig:at_most_jumpramp_hsg}. 
The hybrid base complex represents an abstract (or coarse) version of the input hex-dominant mesh. 
A standard component-based visualization method is to assign distinct colors to different components, as shown in \autoref{fig:at_most_jumpramp_hbc}. The visualization of the hybrid base complex provides an overview of the complexity of a mesh structure. It can also support an initial comparison of the structures between two hex-dominant meshes of the same model. Specifically, the fineness of the structure can be easily perceived via the number of distinct color blocks. The more colored blocks, the more complex the corresponding structure is. 
As in \autoref{fig:compare_twist_cube}, we can easily conclude that the model in \autoref{fig:robust_cube} has a better structure, as it has fewer colored blocks.

While the visualization of the hybrid base complex provides an initial impression of the organization of the blocks along certain directions, the representation of this configuration (or substructure \cite{xu2018hexahedral}) is not direct and requires mental tracing along specific directions. To address this, the visualization of the extracted sheets is needed (\autoref{fig:at_most_jumpramp_sheets}). Before describing sheet visualization, we need to visualize the $VSG$ wireframe, which facilitates the study of individual sheets.

\subsection{Visualize $VSG$ Wireframe}

The purpose of the $VSG$ wireframe is to highlight irregular and non-hex configurations and how they propagate (and the orientation of propagation) throughout the volume. 
To effectively depict this information, we set the opacity values of the individual edges in the resulting $VSG$ wireframe (\autoref{sec:vsgwireframerefinement}) so that the edges at the irregular and non-hex configurations will receive the highest opacity values, while the edges that are further away from the non-hex configurations become more transparent.

To distinguish different structures in a $VSG$ wireframe, we assign different colors to different edges in the $VSG$ wireframe. To reduce the number of colors used while still distinguishing different singularities, a color-coding method is designed based on the connectivity and parallel relationships of edges. 

\begin{definition}[partial parallel singularities]
Given two singularities $s_i$ and $s_j$ consisting of two sets of connected irregular edges with the same valence, respectively. Let $E_i \subset s_i$ and $E_j \subset s_j$ so that $E_i$ and $E_j$ include all edges $e_i \in s_i$ and $e_j\in s_j$ that satisfy $e_i || e_j$ based on \emph{Definition 1}. We say $s_i$ and $s_j$ are partial parallel singularities if $min(\frac{|E_i|}{|se_i|}, \frac{|E_j|}{|se_j|}) >= \rho$, where $\rho$ is a user-specified threshold. 
\end{definition} In our experiments, we set $\rho=0.8$. 

The edges in a pair of partial parallel singularities are assigned the same color.
The two cyan singularities shown in \autoref{fig:at_most_jumpramp_partial_parallel_singularities} are partial parallel singularities. 
Two adjacent non-parallel singular edges will be assigned different colors. This color-coding method provides an effective visualization for $VSG$ wireframe, as shown in \autoref{fig:at_most_jumpramp_VSG}.

\subsection{Visualize Sheets and Their Subsheets}

 \begin{figure}[ht!]
     \centering
     \begin{subfigure}[b]{0.3\columnwidth}
         \centering
         \includegraphics[width=0.99\columnwidth]{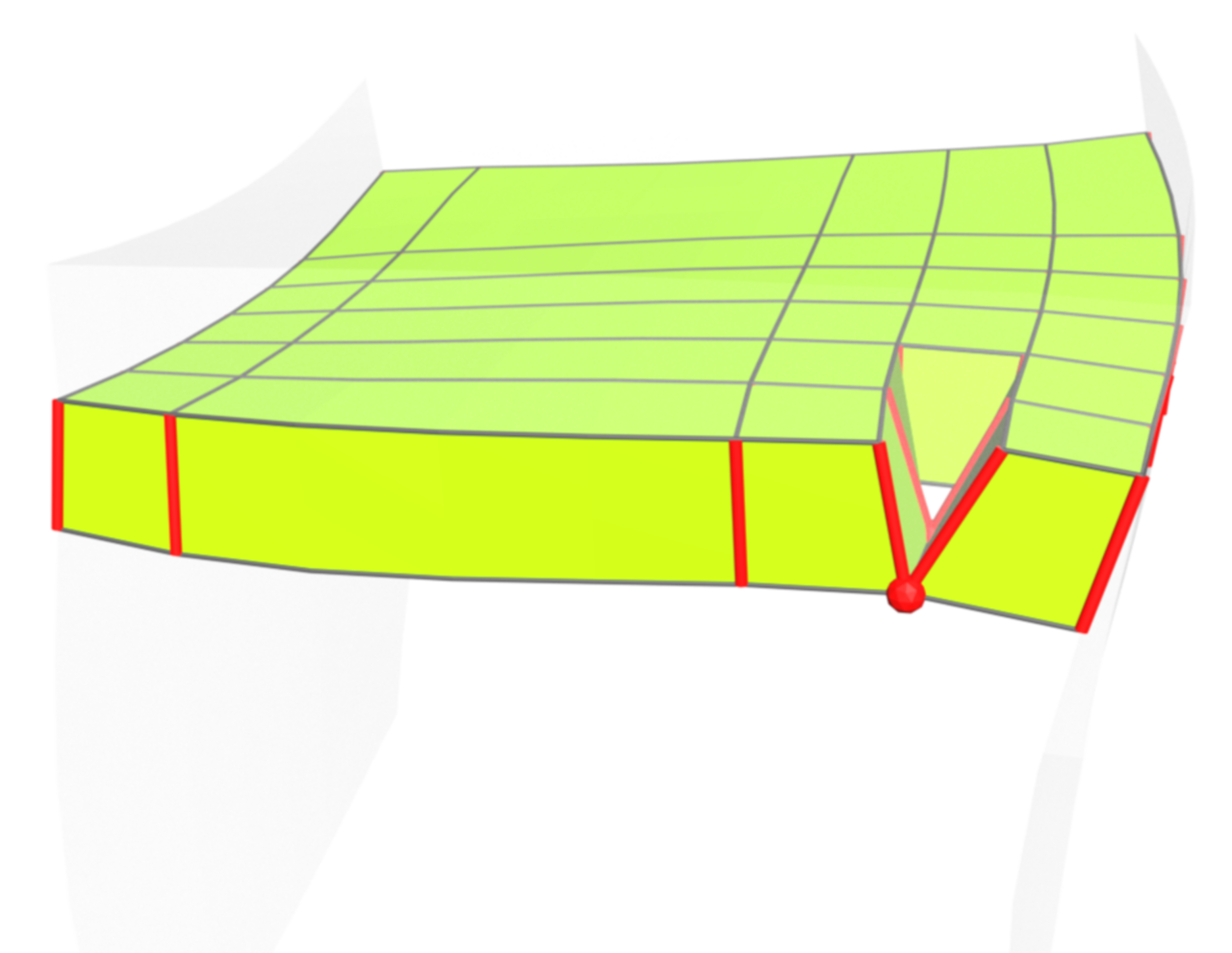}
         \caption{}
         \label{fig:imperfect_type_1}
     \end{subfigure}
     \hspace{0.01\textwidth}
     \begin{subfigure}[b]{0.3\columnwidth}
         \centering
         \includegraphics[width=0.99\columnwidth]{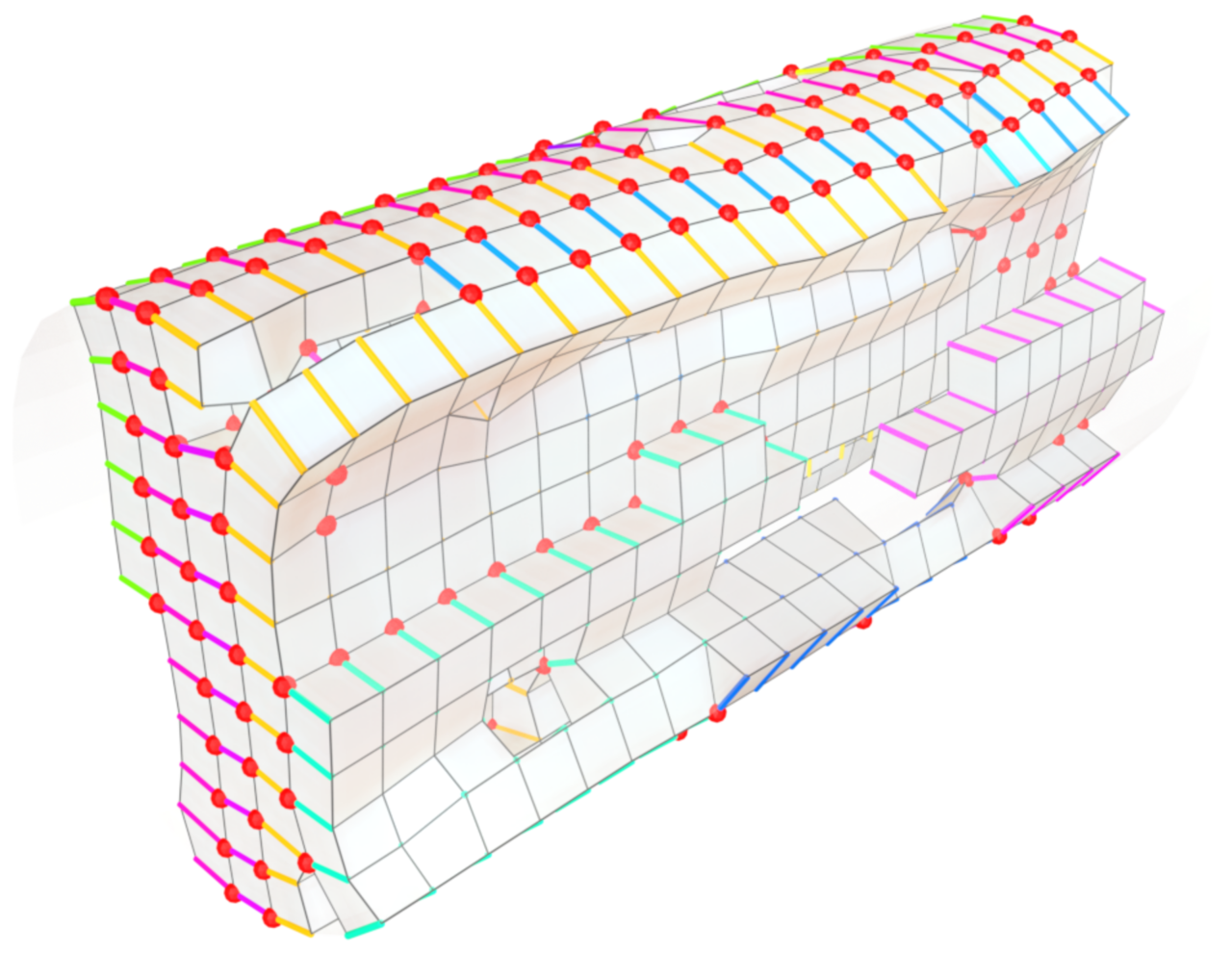}
         \caption{}
         \label{fig:self_parallel_example}
     \end{subfigure}
     \hspace{0.01\textwidth}
     \begin{subfigure}[b]{0.3\columnwidth}
         \centering
         \includegraphics[width=0.99\columnwidth]{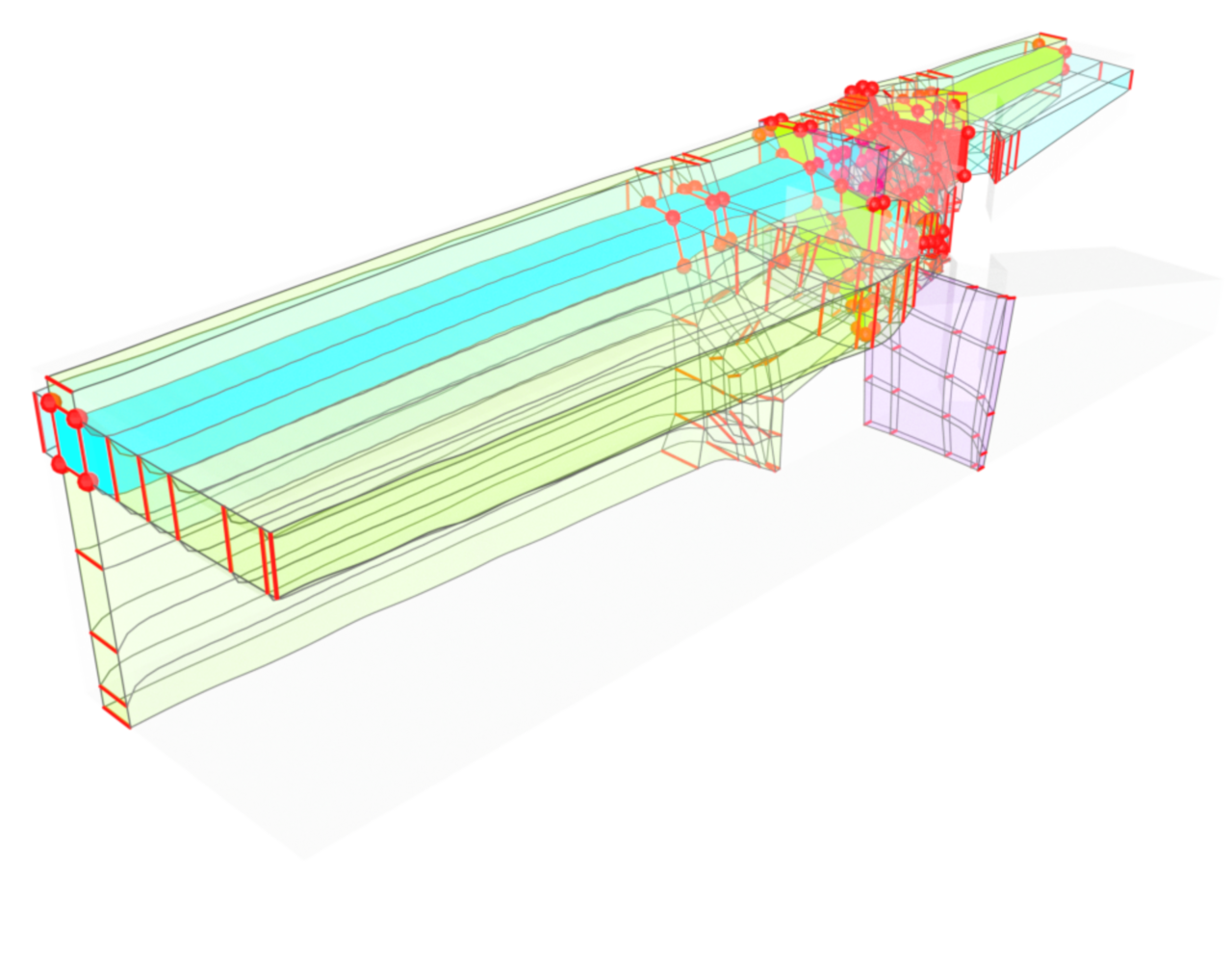}
         \caption{}
         \label{fig:intersect_example}
     \end{subfigure}
    \caption{
    (a) shows an imperfect sheet that contains a few adjacent parallel edges due to the existence of non-hex cells. (b) shows a self-parallel sheet that contains several columns of parallel edges. They are adjacent to each other and oriented in the same direction. (c) shows a self-intersecting sheet that has a few parallel adjacent edges that share the same hex cells, which changes the direction of the sheet. The red dots highlight the unmatched vertices.
    }
    \label{fig:imperfect_sheet_example}
\end{figure}

\begin{figure}[h!]
     \centering
     \includegraphics[width=0.95\columnwidth]{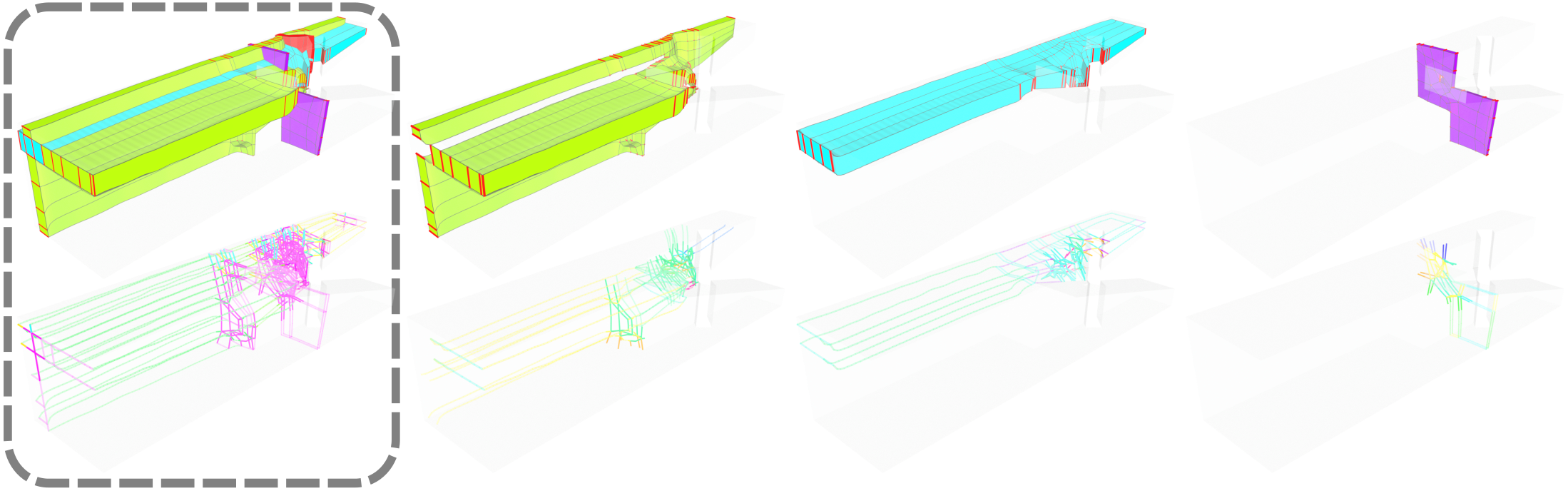}
    \caption{Decomposition of the sheet in \autoref{fig:intersect_example}. This sheet contains 4 subsheets, and the 3 large subsheets are shown. For each subsheet, we also show its corresponding $VSG$ wireframe beneath it.
    }
    \label{fig:sheet_decompose_example}
\end{figure}

We assign a unique color to all components belonging to the same sheet, ensuring that the sheets are easily identifiable, as demonstrated in \autoref{fig:at_most_jumpramp_sheets}. While showing all sheets reveals the possible organizations of the components in the hybrid base complex, 
it easily leads to occlusion. Especially, the imperfect sheets that contribute to the complexity of the structure are often hidden by this visualization. Additional treatment is needed to highlight imperfect sheets.

Recall that the problematic regions of imperfect sheets are usually linked to the unmatched vertices from the above matching process and the self-parallel and self-intersecting configurations of the sheets. We then highlight imperfect sheets through the emphasis on the involved unmatched vertices, self-parallel, and self-intersecting configurations. For self-parallel sheets, we highlight the vertices that are shared by more than one block from the same sheets and the edges that are connected to these vertices but do not share any cell. When we visualize self-intersecting sheets, understanding the regions where cells exhibit self-intersection is very important. Therefore, not only the unmatched vertices should be highlighted, but the cells that contain neighboring parallel edges in a sheet should also be displayed, which is crucial for a comprehensive global analysis. Examples of these visualizations can be found in \autoref{fig:imperfect_sheet_example}.

The self-intersecting sheets can be further decomposed into a number of subsheets for a detailed analysis. In this case, we visualize the individual subsheets as above for the sheets. In addition, we visualize the $VSG$ wireframe within each subsheets to describe the internal configurations of the subsheet. See \autoref{fig:sheet_decompose_example} for some examples of this visualization. 

\section{Results}
\label{sec:results}
 
 We apply the hybrid base complex construction and its visualization to analyze and compare three sets of hex-dominant meshes. 
 The first set is produced by At-Most-Hexa meshing \cite{bukenberger2022most}, which we downloaded from hexalab.net \cite{BRACCI201924} and contains 7 relatively simple models. The second set is released by the Robust-Hex-Dominant meshing \cite{gao2017robust}, which contains 106 models, including many models with high genus and sharp features.  The third set is from LoopyCuts \cite{livesu2020loopycuts}, which has 39 models. To our knowledge, those models are all we can obtain from current state-of-the-art hex-dominant meshing methods.

\subsection{Structural Analysis of Hex-dominant Meshes}

\autoref{fig:teaser} demonstrates the process of using our hybrid base complex and its decomposition and visualization to support the analysis of the structural configurations of a cylinder hex-dominant mesh produced by the At-Most-Hexa method \cite{bukenberger2022most}. This mesh has 1655 3D cells. 1545 of them are hexes  (i.e., 93.3\% of cells are hexes). 
(a)--(c) show the steps of constructing the hybrid base complex, and (d)--(f) show the visual analysis of an imperfect sheet in the obtained hybrid base complex. The direct visualization (a) that most hex-dominant meshing techniques use shows a pure hex configuration on the boundary of the cylinder and a few simple non-hex cells (e.g., triangles) on the base face. 
However, our hybrid singularity graph (b) reveals that the surface non-hex cells propagate through the volume. While these non-hex cells only stack along the direction of the cylinder, they result in a complex hybrid base complex (i.e., with many components, as shown by different colored blocks in (c)). 
Next, we inspect an imperfect sheet (d) that occupies a larger volume of the cylinder. This sheet has a complex boundary as shown by its corresponding $VSG$ (e). 
(f) shows the simplified $VSG$ wireframe derived from $VSG$. It aims to show the orientation and propagation of the irregular and non-hex configurations. It has a dense configuration. 

\begin{figure}[t!]
    \centering
        \begin{subfigure}[b]{0.42\columnwidth}
        \centering
        \includegraphics[width=1.0\columnwidth]{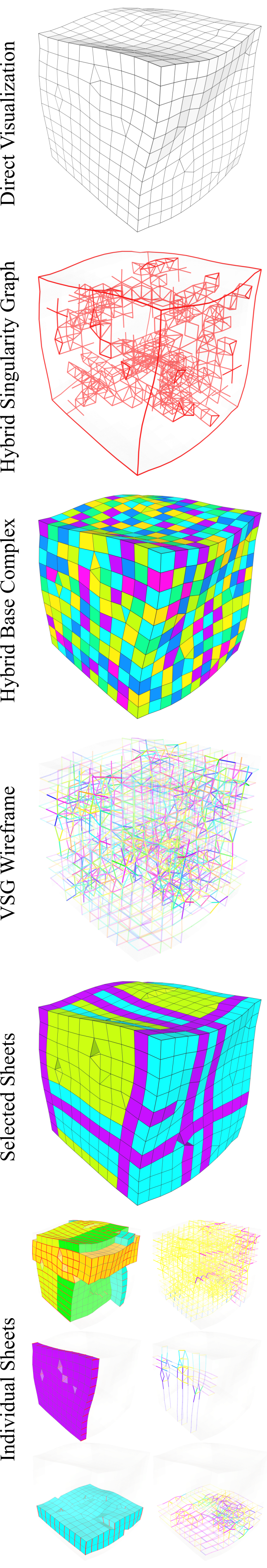}
        \caption{}
        \label{fig:at_most_cube}
        \end{subfigure}
        \begin{subfigure}[b]{0.42\columnwidth}
        \centering
        \includegraphics[width=1.0\columnwidth]{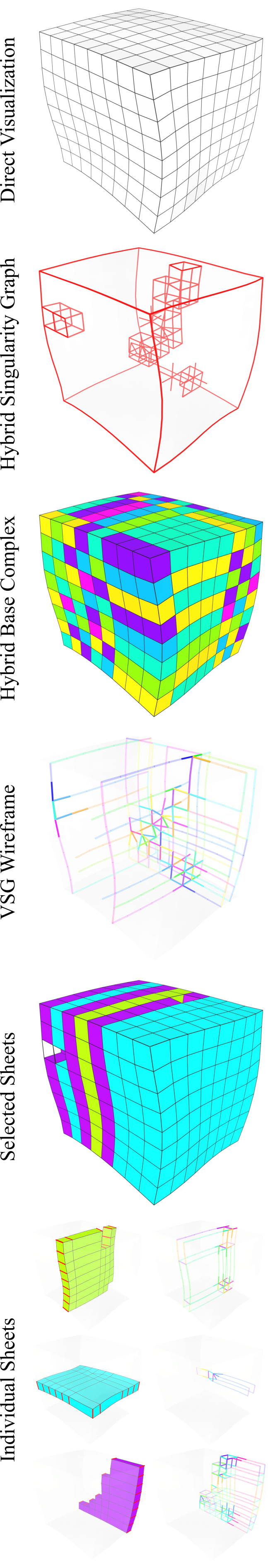}
        \caption{}
        \label{fig:robust_cube}
    \end{subfigure}
    \caption{Comparison of the structural configurations of the hex-dominant meshes of a twisted cube produced by the At-Most-Hexa method (a) and the Robust Hex-Dominant meshing (b), respectively.
    }
    \label{fig:compare_twist_cube}
\end{figure}

To decipher the detailed configurations, we further decompose this imperfect sheet into a few subsheets (\autoref{sec:sheetdecomp}) (g). Each subsheet is displayed with a unique color and is defined by a set of parallel edges (i.e., the thick red edges). For each subsheet, we construct and visualize its $VSG$ wireframe. 
We now can see the orientation of the irregular and non-hex configurations clearly using the $VSG$ wireframes. A close inspection reveals that, while the yellow subsheet has a small number of non-hex cells, they are oriented in different directions (e.g., the opening of the ``Y'' shapes in the upper row is different from the one in the lower row). However, these two differently oriented configurations can be removed separately using the strategy demonstrated in \autoref{fig:valence_based_sg}g--h, as they do not seem to overlap (or intersect) with each other. When inspecting the purple subsheet, we notice that non-hex configurations are oriented consistently in the vertical portion of the subsheet, but differently in the horizontal part. This indicates that the former can be easily removed, while the latter may not. Other subsheets and their $VSG$ wireframes can be analyzed similarly.
Please zoom in on the $VSG$ wireframe images to see the above details. 
A similar multi-level study can be performed on the two twisted cube meshes shown in \autoref{fig:compare_twist_cube}, which will be described later.

\autoref{fig:sheet_decompose_example} provides a detailed analysis of an imperfect sheet of \autoref{fig:intersect_example} from the $scarf3$ hex-dominant mesh.
This sheet (in the dashed box) contains self-intersecting configurations, in which the parallel edges point to different directions caused by the direction change at the intersections. Its $VSG$ wireframe shows many non-hex configurations (with ``Y''-shape configuration) that form a cluster, which prevents an effective study of how they impact the structure of the sheet. 
We then decompose the sheet into 4 subsheets with only type-1 or type-2 configuration. The 3 largest subsheets are shown. While 2 subsheets in the right columns have simple plane-like configurations (so are their respective $VSG$ wireframes), the subsheet in the second column contains two separated groups of non-hex configurations oriented in two different directions. Among them, the group in the middle part of this subsheet has non-hex configurations with consistent orientation, suggesting a possibility of removing all of them.

 \subsection{Compare Hex-dominant Meshes Produced by Different Methods}
 \label{sec:comparison}

We next apply our method to study the differences in the structure characteristics of the hex-dominant meshes produced by different methods, which was hard to achieve previously.

\begin{figure*}[ht!]
    \centering
        \begin{subfigure}[b]{\linewidth}
        \centering
        \includegraphics[width=\columnwidth]{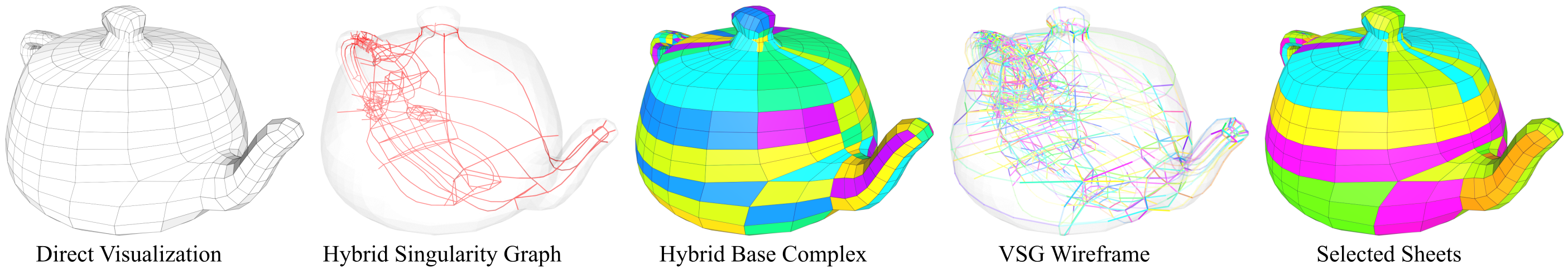}
        \caption{LoopyCuts}
        \label{fig:loopy_teapot}
        \end{subfigure}
        \begin{subfigure}[b]{\linewidth}
        \centering
        \includegraphics[width=\columnwidth]{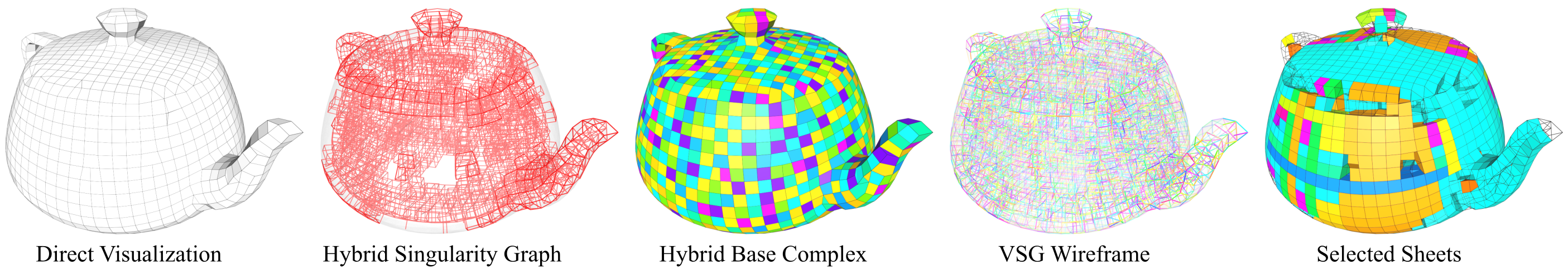}
        \caption{Robust hex-dominant meshing}
        \label{fig:robust_teapot}
    \end{subfigure}
    \caption{Comparison of the structural configurations of the hex-dominant meshes of a teapot model produced by the LoopyCuts method (a) and the Robust-Hex-Dominant Meshing (b), respectively. 
    }
    \label{fig:compare_robust_loopy_teapot}
\end{figure*}

In \autoref{fig:compare_twist_cube}, we compare the results of two different algorithms, i.e., the At-Most-Hexa Meshing (a) and the Robust-Hex-Dominant Meshing (b), on the Cube Twist model. It is obvious that At-Most-Hexa produces a mesh with a large number of non-hex elements, resulting in a very complex singularity graph (second row) as well as many small components in the obtained hybrid base complex (third row). In contrast, the Robust-Hex-Dominant places non-hex elements in fewer places, resulting in a simpler singularity graph and a hybrid base complex with fewer and larger blocks. Additionally, the extracted sheets  (fifth row) show that the mesh by the Robust-Hex-Dominant has larger and simpler (e.g., plane-like) sheets (with fewer holes or gaps in the sheets), indicating the simplicity of its structure. To study the complexity of the irregular and non-hex configurations, we compare their respective $VSG$ wireframes (fourth row). Visually, the $VSG$ wireframe of the mesh by the Robust-Hex-Dominant is much cleaner than the one by the At-Most-Hexa. Not only it has fewer irregular and non-hex configurations, but also their orientations are aligned with one of the three principal directions. This is probably because the Robust-Hex-Dominant is a field-aligned method, in which the octohedral field used to guide the placement of 3D cells is aligned with the three principal directions of the cube. 

The last row of \autoref{fig:compare_twist_cube} shows some representative imperfect sheets from the two obtained hybrid base complexes, respectively. Again, the sheets from the mesh by the Robust-Hex-Dominant are simpler than those from the other mesh. Specifically, they are plane-like and aligned with the surface features. This simplicity is also reflected by their respective $VSG$ wireframes, each of which contains similarly oriented irregular and non-hex configurations that can be removed. In contrast, the mesh by the At-Most-Hexa has a complex imperfect sheet that possesses self-intersecting configurations (i.e., the first sheet shown in the last row). This imperfect sheet can be decomposed into several simpler subsheets (shown by different colored blocks) that are mutually orthogonal to each other. The complex configuration of this imperfect sheet is also reflected by its $VSG$ wireframe which exhibits a dense configuration. The (cyan) sheet shown at the bottom of (a) has a non-planar configuration (i.e., not all involved hexes are on the same plane). This is caused by the presence of a few non-hex elements that lift their adjacent hex cells. While these non-hex cells are simple individually (i.e., triangle-like), they have different orientations and intersect with each other, as shown by the $VSG$ wireframe, suggesting that removing them may not be easy. 

\begin{figure*}[ht!]
    \centering
        \begin{subfigure}[b]{\linewidth}
        \centering
        \includegraphics[width=\columnwidth]{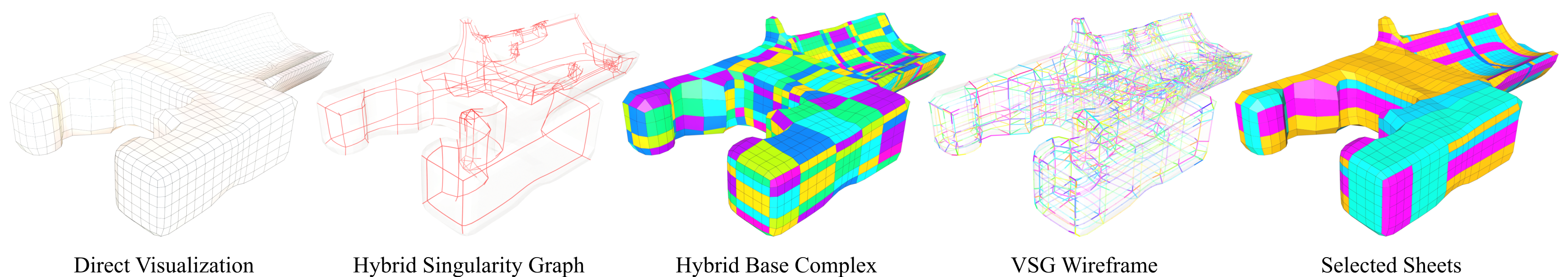}
        \caption{LoopyCuts}
        \label{fig:loopy_fertility}
        \end{subfigure}
        \begin{subfigure}[b]{\linewidth}
        \centering
        \includegraphics[width=\columnwidth]{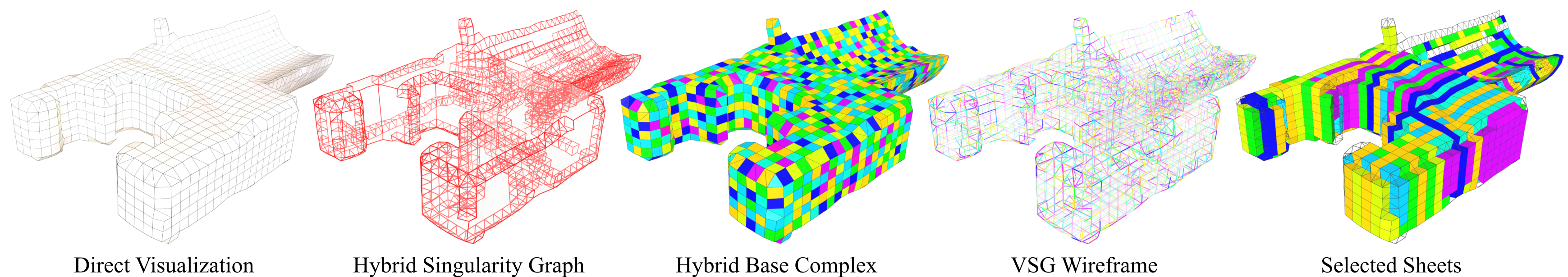}
        \caption{Robust hex-dominant meshing}
        \label{fig:robust_fertility}
    \end{subfigure}
    \caption{Comparison of the structural configurations of the hex-dominant meshes of a blade model produced by the LoopyCuts method (a) and the Robust-Hex-Dominant Meshing (b), respectively. 
    }
    \label{fig:compare_robust_loopy_blade}
\end{figure*}

A comparison between the LoopyCuts and the Robust-Hex-dominant method is shown in \autoref{fig:compare_robust_loopy_blade} and \autoref{fig:compare_robust_loopy_teapot}, respectively. It is apparent that LoopyCuts takes advantage of simple cuts, resulting in  a simpler singularity graph (second column) and a hybrid base complex with larger blocks (third column). However, due to the inherent limitations of loops and the emphasis on a smaller number of cuts, the structure may exhibit large distortion in certain areas, e.g., the top of the teapot and the far end of the blade where the sizes of the components shrink (last column). In contrast, the Robust-Hex-Dominant method produces a more regular structure (i.e., with uniform-size components). 
In the meantime, while both methods place non-hex elements along the feature lines of the blade model (see their respective $VSG$ wireframes) and the orientation of these non-hex elements are aligned with the principal directions of the model, they behave significantly differently on the teapot model. In particular, the irregular and non-hex elements are uniformly distributed in the mesh by the Robust-Hex-Dominant, illustrated by the evenly distributed color lines in the $VSG$ wireframe. In contrast, those non-hex elements are placed and oriented irregularly in the mesh by the LoopyCuts. This will make their future removal challenging. The different behavior of LoopyCuts on the two models suggests that LoopyCuts may produce better meshes for CAD models where their sharp features can provide guidance to the cutting, while it may not work well for organic-like models where the intrinsic cuts are hard to derive. Nevertheless, LoopyCuts still produces hex-dominant meshes with fewer non-hex elements than the Robust-Hex-Dominant meshing.

The above examples demonstrate that the proposed structure facilitates the characterization of the global (or structural) properties of individual hex-dominant meshes and the comparison of structure differences between two meshes, which was previously impossible. 
By using hybrid base complexes and their respective substructures, we can observe how a few non-hex elements can result in intricate configurations in the extracted hybrid base complex, particularly when these elements are located in the model's interior. For example, the teapot mesh produced by LoopyCuts (\autoref{fig:loopy_teapot}) contains only 24 non-hex cells (out of 2404 cells), and all these non-hex cells are simple with at most 6 faces each. However, they result in a complex interior structure as shown by the $VSG$ wireframe.
While traditional visualizations may indicate the locations of non-hex cells locally, they cannot depict how these non-hex cells affect the mesh configurations globally. Our multi-level structure representation addresses this issue. 

\subsection{Performance}

Our hybrid base complex construction and the derivation of other substructure representations are implemented using Python. The proposed visualizations are achieved using Blender via customized Python plugins. 
We have applied our implementation to the three sets of hex-dominant meshes without failure. \autoref{tab:statistics} provides the statistics for the meshes used in the paper. All timing information was measured on a workstation running Windows 11 and with an Intel 12th i5 CPU, Navida RTX 2070 GPU and 32 GB RAM. The statistics for all meshes that we have experimented with can be found in the supplemental materials.

\begin{table*}[!ht]
\centering
\begin{tabular}{c|cc|cc|cccc|c|ccccc} 
Models               & $|C|$ & $\frac{|H|}{|C|}$ & $|C_B|$ & $\frac{|H_B|}{|C_B|}$ & $|G_{SH}|$ & $|G_{SH}^{T1}|$ & $|G_{SH}^{T2}|$ & $|G_{SH}^{T3}|$ & $|G_{SH*}|$ & $T_{G_S}$ & $T_{G_B}$ & $T_{VSGW}$ & $T_{G_{SH}}$ & $T_{G_{SH*}}$ \\ \hline
$blade^{\dagger}$    & 4125  & 0.994             & 1107    & 0.98                  & 34       & 3               & 0               & 1               & 23          & 0.32      & 7.33      & 6.39       & 0.25       & 0.32          \\
$blade^{\ddag}$      & 2751  & 0.774             & 2751    & 0.77                  & 171      & 16              & 2               & 1               & 2           & 0.62      & 7.42      & 36.4      & 0.81       & 0.22          \\
$cylinder_g^*$       & 1655  & 0.934             & 1275    & 0.91                  & 28       & 19              & 0               & 1               & 8           & 0.14      & 2.57      & 2.54       & 0.45       & 0.09          \\
$cylinder_n^*$       & 1671  & 0.858             & 1671    & 0.86                  & 25       & 15              & 1               & 1               & 30          & 0.19      & 3.02      & 11.32      & 0.41       & 0.79          \\
$scarf 3^{\dagger}$  & 5877  & 0.996             & 763     & 0.97                  & 36       & 6               & 1               & 3               & 4           & 0.44      & 13.64     & 3.13          & 0.16       & 0.02          \\
$jumpRamp^*$         & 1460  & 0.973             & 352     & 0.89                  & 26       & 0               & 1               & 0               & 0           & 0.11      & 1.18      & 0.28        & 0.08       & 0             \\
$teapot^\dagger$     & 2404  & 0.99              & 772     & 0.97                  & 23       & 5               & 1               & 2               & 3           & 0.18      & 2.88      & 2.66       & 0.23       & 0.01          \\
$teaport^{\ddag}$    & 5844  & 0.797             & 5844    & 0.8                   & 124      & 20              & 12              & 1               & 0           & 2.15      & 24.21     & 199.47     & 3.32       & 0             \\
$twistcube\_s^*$     & 1301  & 0.889             & 1301    & 0.89                  & 20       & 16              & 2               & 1               & 9           & 0.12      & 2.09      & 6.13       & 0.36       & 0.3           \\
$twist cube^{\ddag}$ & 493   & 0.963             & 221     & 0.92                  & 19       & 8               & 0               & 0               & 0           & 0.04      & 0.33      & 0.16       & 0.03       & 0            
\end{tabular}

\caption{$\dagger$ indicate meshes from LoopyCuts, $\ddag$ Robust-Hex-Dominant mesh, and $*$ At-Most-Hexa.  $|C|$ indicates the number of cells, $\frac{|H|}{|C|}$ the hex ratio, $|C_B|$ the number of hybrid base complex components, $\frac{|H_B|}{|C_B|}$ the ratio of hex base complex components, $|G_{SH}|$ the number of extracted sheets, $|G_{SH}^{T1}|$ the number of type-1 imperfect sheet, $|G_{SH}^{T2}|$ the number of self-parallel imperfect sheets, $|G_{SH}^{T3}|$ the number of self-intersecting sheets, $|G_{SH*}|$ the number of subsheets for the largest self-intersecting sheet, $T_{G_S}$ the time to extract hybrid singularity graph, $T_{G_B}$ the time to extract hybrid base complex, $T_{VSGW}$ the time to extract a VSG wireframe of the mesh, $T_{G_{SH}}$ the time to extract sheets, and $T_{G_{SH*}}$ the time to extract all sub-sheet for the largest self-intersecting sheet.}
\label{tab:statistics}
\end{table*}

\section{Conclusion and Future Work}

This paper presents the first structure, called the hybrid base complex, for hex-dominant meshes. This structure covers both hexahedral elements and non-hexahedral elements. Our structure is a natural extension of the conventional base complex for all-hex meshes. With this structure, we can evaluate the quality of the hex-dominant meshes in a more global sense which was impossible previously. 
To support an effective study of the structure of a hex-dominant mesh, especially the configurations of the non-hex elements and their impact on the subsequent simplification process, we extract a few substructures of the structure, including sheets, valence-based singularity graph $VSG$, and $VSG$ wireframe. These substructures reduce the visual complexity and support a multi-level study of the structure configurations of a hex-dominant mesh. To further reduce the clutter in the visualization of the extracted substructures, a number of simplification strategies are introduced to further decompose a complex sheet into simpler subsheets and to reduce the geometric elements in the $VSG$ wireframe. The extraction of the hybrid base complex and its substructures leads to an integrated framework for visual analysis of the structures of various hex-dominant meshes. We have applied our framework to over 120 hex-dominant meshes produced by three state-of-the-art meshing techniques to analyze their structural characteristics and compare their structural differences. The results demonstrate the effectiveness of our method. 

Several future directions can be explored based on our work. First, the proposed structure representations (including the hybrid singularity graph, hybrid base complex, and imperfect sheets) can be used to define comprehensive metrics to quantify the structural complexity of a given hex-dominant mesh in a similar fashion to the work \cite{xu2018hexahedral}. Second, the imperfect sheet decomposition and the construction of the $VSG$ wireframe set the foundation for the development of effective operations to procedurally remove certain non-hex elements and improve the hex-dominant mesh quality. Third, the current visualization of $VSG$ wireframe does not differentiate non-hex configurations of different types (e.g., with different numbers of faces), which can be improved. Finally, the proposed analysis and visualization framework based on the proposed structural representations can be integrated into a stand-alone or web-based visualization system, making it more accessible to the meshing community. Finally, the extraction of some structural representations can be optimized to achieve real-time performance.

\ifCLASSOPTIONcompsoc
  \section*{Acknowledgments}
\else
  \section*{Acknowledgment}
\fi

We would like to thank the anonymous reviewers for their valuable feedback and comments to this work.
This research was supported by NSF OAC 2102761.

\ifCLASSOPTIONcaptionsoff
  \newpage
\fi

\bibliographystyle{IEEEtran}
\bibliography{Contents/paper-bibliography}




%

\begin{IEEEbiography}[{\includegraphics[width=1in,height=1.25in,clip,keepaspectratio]{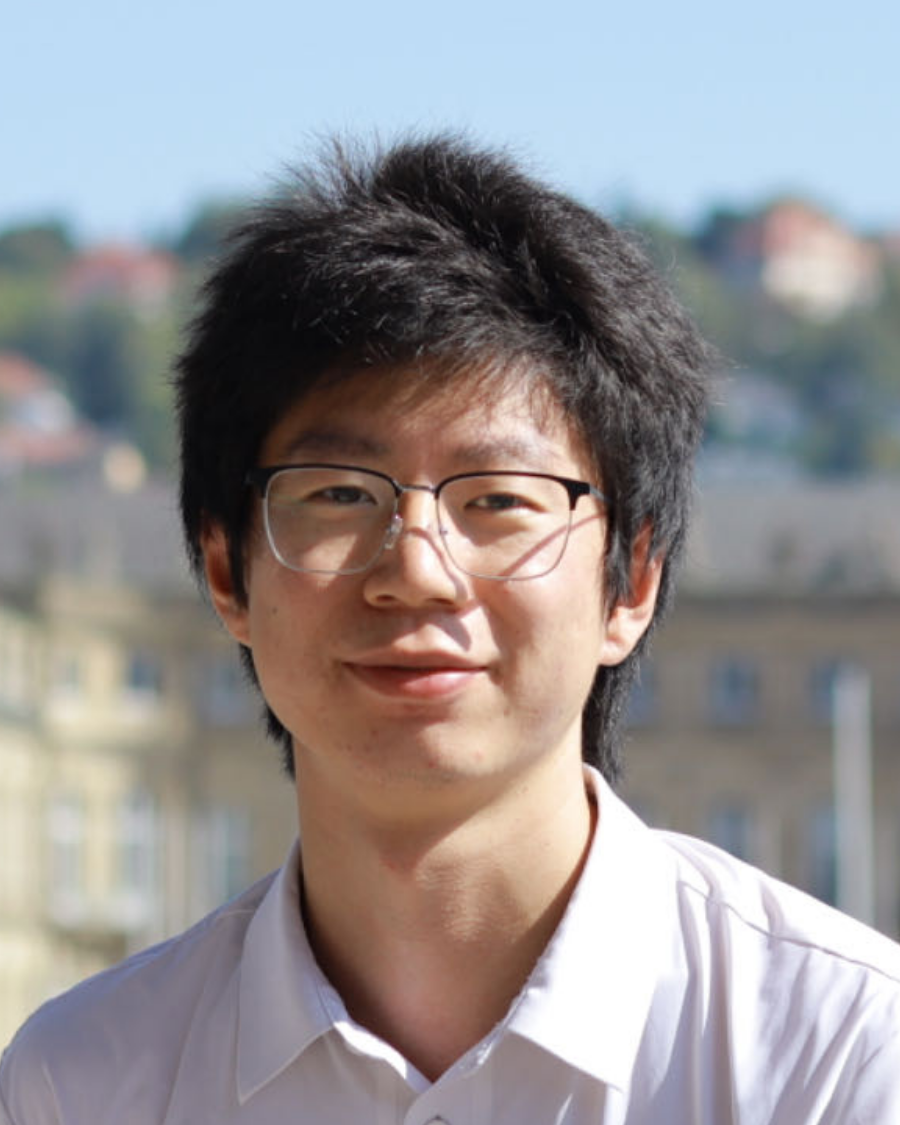}}]
{Lei Si}
is a Ph.D. student in Computer Science from the Department of Computer Science, University of Houston. He earned an M.S. degree with honors in Computer Science from the University of Illinois at Springfield in 2020, And B.E. in Cybersecurity at North China University of Technology in 2018, His research focuses on geometric modeling, visualization, physically-based simulation, computer vision, cyber security, and artificial intelligence.
\end{IEEEbiography}

\begin{IEEEbiography}[{\includegraphics[width=1in,height=1.25in,clip,keepaspectratio]{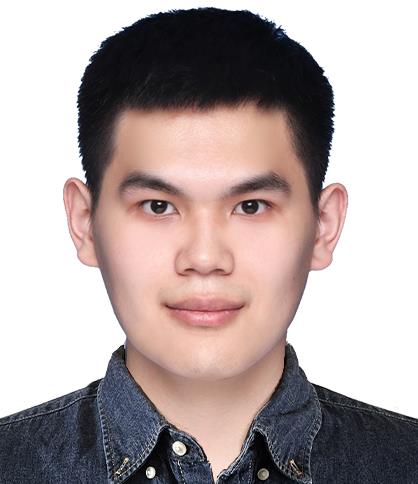}}]
{Haowei Cao}
is a Master of Science candidate in Electrical and Computer Engineering at Johns Hopkins University. His prior studies at the University of Houston centered on Ph.D.-level research in Computer Science, delving into Image Processing, Computer Vision, and Artificial Intelligence. he achieved his Bachelor of Science in Computer Science, minoring in Mathematics, from the University of Tennessee-Knoxville, graduating Summa Cum Laude.
\end{IEEEbiography}

\begin{IEEEbiography}[{\includegraphics[width=1in,height=1.25in,clip,keepaspectratio]{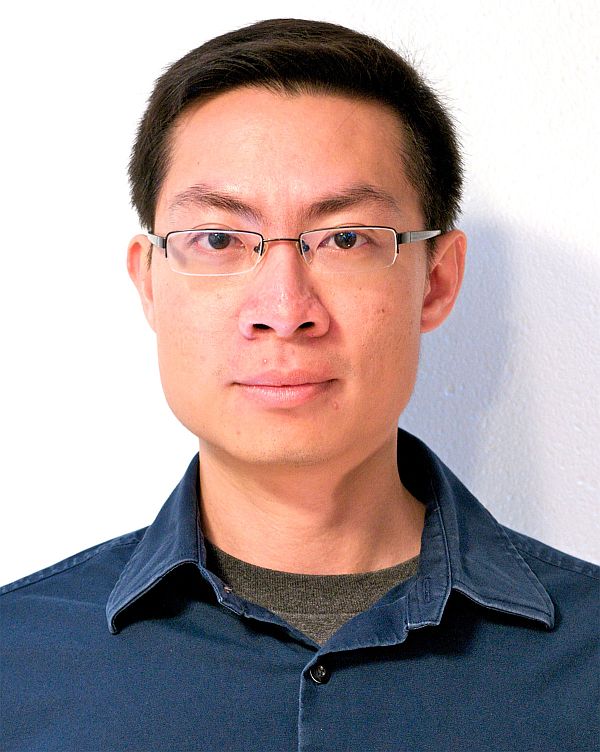}}]{Guoning Chen}
is an Associate Professor at the Department of Computer Science at the University of Houston. He earned a Ph.D. degree in Computer Science from Oregon State University in 2009. His research interests include visualization,
data analytics, computational topology, geometric modeling, geometry processing and physically-based simulation. He is a senior member of IEEE and a member of ACM.
\end{IEEEbiography}








\end{document}